\RequirePackage{silence}
\documentclass[%
 reprint,
superscriptaddress,
nofootinbib,
 amsmath,amssymb,
 aps,
pra,
floatfix,
]{revtex4-2}

\usepackage[T1]{fontenc}    
\usepackage{lmodern}
\usepackage[english]{babel} 
\usepackage[table,dvipsnames]{xcolor}
\usepackage[colorlinks=true,
			hyperindex, 
		    linkcolor = NavyBlue,
		    anchorcolor = hyperrefblue,
		    citecolor = Green,
		    filecolor = hyperrefblue,
		    urlcolor = NavyBlue,
			breaklinks=true]{hyperref}

\usepackage{amsthm}

\usepackage[capitalize]{cleveref} 
\crefname{figure}{Figure}{Figures} 

\usepackage{paralist}
\usepackage{enumitem}

\usepackage{url}            
\usepackage{booktabs}       

\usepackage{amsmath,amssymb,mathtools,dsfont} 
\usepackage{pifont}
\usepackage{etoolbox}

\usepackage{complexity}
\let\PP\undefined 

\usepackage{microtype}      
\usepackage{multirow}       
\usepackage{algorithm}
\usepackage{algpseudocode}
\algrenewcommand\algorithmiccomment[1]{\hfill $\triangleright$ #1} 
\usepackage[version=4]{mhchem}         
\usepackage{graphicx}
\usepackage{grffile}
\usepackage{xcolor}
\usepackage{tikz}
\usepackage{pgfplots}
\DeclareUnicodeCharacter{2212}{−}
\usepgfplotslibrary{groupplots,dateplot}
\usetikzlibrary{patterns,shapes.arrows}
\pgfplotsset{compat=newest}
\usepackage{tikzscale}
\usepackage{rotating}
\usepackage{amsmath}
\usepackage{braket}
\usepackage{placeins}

\definecolor{ShadowGrouping}{HTML}{BA2D0B}
\definecolor{Adaptive}{HTML}{0A8754}
\definecolor{Derand}{HTML}{C08497}
\definecolor{AEQUO}{HTML}{EFC88B} 
\definecolor{Random}{HTML}{01110A}
\definecolor{Overlapped}{HTML}{1098F7}
\definecolor{Guarantee}{RGB}{102,102,102}

\usepackage[
nolist]{acronym}
\makeatletter
\AtBeginDocument{%
  \renewcommand*{\AC@hyperlink}[2]{%
    \begingroup
      \hypersetup{hidelinks}%
      \hyperlink{#1}{#2}%
    \endgroup
  }%
}
\makeatother

\renewcommand{\Re}{\operatorname{Re}}
\renewcommand{\Im}{\operatorname{Im}}

\newcommand{\fro}{\mathrm{F}}

\NewDocumentCommand\Cl{mg}{
    \ensuremath{\mathrm{Cl}_{#1}\IfNoValueTF{#2}{}{(#2)}}%
}
\NewDocumentCommand\HW{mg}{
    \ensuremath{\mathrm{HW}_{#1}\IfNoValueTF{#2}{}{(#2)}}%
}

\newcommand{\PP}{\mathbb{P}}

\DeclarePairedDelimiterX{\abs}[1]{\lvert}{\rvert}{%
  \ifblank{#1}{\,\cdot\,}{#1}
}   

\DeclarePairedDelimiterX\norm[1]\lVert\rVert{%
  \ifblank{#1}{\,\cdot\,}{#1}
}   

\DeclarePairedDelimiterX{\iiiNorm}[1]{\lvert}{\rvert}{%
  \delimsize\lvert\delimsize\lvert#1\delimsize\rvert\delimsize\rvert%
}

\DeclarePairedDelimiterXPP\snorm[1]{}\lVert\rVert{_\infty}{\ifblank{#1}{\,\cdot\,}{#1}}   

\DeclarePairedDelimiterXPP\twonorm[1]{}\lVert\rVert{_2}{\ifblank{#1}{\,\cdot\,}{#1}}   

\DeclarePairedDelimiterXPP\trnorm[1]{}\lVert\rVert{_1}{\ifblank{#1}{\,\cdot\,}{#1}}   

\DeclarePairedDelimiterXPP\fnorm[1]{}\lVert\rVert{_{\fro}}{\ifblank{#1}{\,\cdot\,}{#1}}   

\DeclarePairedDelimiterXPP\dnorm[1]{}\lVert\rVert{_\diamond}{\ifblank{#1}{\,\cdot\,}{#1}}   

\DeclarePairedDelimiterXPP\cbnorm[1]{}\lVert\rVert{_\mathrm{cb}}{\ifblank{#1}{\,\cdot\,}{#1}}   
\DeclarePairedDelimiterXPP\onenorm[1]{}\lVert\rVert{_{1\rightarrow 1}}{\ifblank{#1}{\,\cdot\,}{#1}}   
\DeclarePairedDelimiterXPP\ddnorm[1]{}\lVert\rVert{_{\diamond\rightarrow \diamond}}{\ifblank{#1}{\,\cdot\,}{#1}}   
\DeclarePairedDelimiterXPP\ssnorm[1]{}\lVert\rVert{_{\infty\rightarrow\infty}}{\ifblank{#1}{\,\cdot\,}{#1}}   

\let\Set\relax
\DeclarePairedDelimiterX\Set[1]\{\}{%
  
  #1
}

\DeclarePairedDelimiterX\innerp[2]{\langle}{\rangle}{%
  \ifblank{#1}{\,\cdot\,}{#1} , \ifblank{#2}{\,\cdot\,}{#2}%
}

\DeclarePairedDelimiterX\average[1]{\langle}{\rangle}{%
  \ifblank{#1}{\,\cdot\,}{#1}%
}

\providecommand{\bra}[1]{\left\langle #1 \right\vert}
\providecommand{\ket}[1]{\left\vert #1 \right\rangle}
\providecommand{\braket}[1]{\left\langle #1 \right\rangle}

\DeclarePairedDelimiterX\sandwich[3]{\langle}{\rangle}%
  {#1\,\delimsize\vert\kern0.15ex\mathopen{}#2\kern0.15ex\delimsize\vert\kern0.15ex\mathopen{}#3}

\DeclarePairedDelimiterX\obraket[2]{(}{)}%
  {#1\kern0.15ex\delimsize\vert\kern0.15ex\mathopen{}#2}

\DeclarePairedDelimiterX\oketbra[2]{\vert}{\vert}%
  {#1\kern0.15ex\delimsize)\delimsize(\kern0.15ex\mathopen{}#2}

\DeclarePairedDelimiterX\osandwich[3]{(}{)}%
  {#1\,\delimsize\vert\kern0.15ex\mathopen{}#2\kern0.15ex\delimsize\vert\kern0.15ex\mathopen{}#3}

\renewcommand{\Pr}{\operatorname{\PP}}

\begin{document}
\title{CANOE: Classically Assisted Non-Orthogonal Eigensolver}

\author{Jihyeon Park}
\email{park687@wisc.edu}
\affiliation{%
    Department of Physics,
    University of Wisconsin -- Madison,
    Madison, WI 53706, USA
}%
\author{Collin C. D. Frink}%
\affiliation{%
    Department of Physics,
    University of Wisconsin -- Madison,
    Madison, WI 53706, USA
}%
\author{Matthew Otten}%
\email{mjotten@wisc.edu}
\affiliation{%
    Department of Physics,
    University of Wisconsin -- Madison,
    Madison, WI 53706, USA
}%
 \affiliation{Department of Chemistry, University of Wisconsin -- Madison, Madison, WI 53706, USA}

\begin{abstract}
In the early fault-tolerant regime, where quantum resources remain limited, hybrid quantum-classical 
strategies offer one possible route toward quantum advantage. We introduce CANOE, the Classically Assisted 
Non-Orthogonal Eigensolver, as such an approach, distributing Rayleigh–Ritz basis states between quantum 
and classical hardware. This approach leverages the expressive power of quantum states, which are costly 
to reproduce classically, while augmenting them with a large pool of classically generated basis states 
that can be incorporated at negligible computational cost. We validate this through numerical simulations 
of a 76-qubit chromium atom system, quantifying how each additional quantum basis state enhances 
ground-state representability and how the inclusion of classical states further amplifies this 
improvement. Such a hybrid basis framework necessarily requires an efficient protocol on 
quantum hardware for evaluating overlaps between quantum and classical states in the resulting 
generalized eigenvalue formulation. We address this by introducing a histogram-based protocol 
and demonstrate through numerical simulations that it can approach chemical accuracy at moderate 
sampling cost. To solve the resulting generalized eigenvalue problem stably, CANOE incorporates a 
Schur-complement-based stabilization procedure that mitigates ill-conditioning caused by linear 
dependencies in the hybrid basis. Taken together, these results position CANOE as a practical 
framework for combining limited quantum resources with expansive classical resources for
early fault-tolerant quantum simulations.
\end{abstract}

\maketitle
\hypersetup{
pdftitle={CANOE: Classically Assisted Non-Orthogonal Eigensolver},
pdfsubject={Quantum many-body physics},
pdfauthor={Jihyeon Park, Collin Frink, Matthew Otten},
pdfkeywords={
				}
}

\section{Introduction}
\label{sec:introduction}
%

Quantum computers have the potential to provide distinct computational advantage
for problems in chemistry and materials science~\cite{alexeev2025perspective,alexeev2024quantum}, especially with 
large-scale, fault-tolerant devices~\cite{gratsea2025achieving}. The pursuit of quantum advantage in quantum 
simulation remains an active effort in the early fault-tolerant regime~\cite{Lanyon2010, Peruzzo2014, Tilly2022}. 
Despite steady progress in hardware development, quantum devices continue to operate under 
constraints in coherence, gate fidelity, and feasible circuit depth~\cite{Mohseni2024}. Consequently, hybrid quantum–
classical frameworks have emerged as a practical strategy for
distributing computational workload between quantum and classical processors, thereby optimizing the use of scarce 
quantum resources~\cite{McClean2017QSE, Tilly2022, Hao2026Geometry}. The most representative example of such hybrid 
approaches is the variational quantum eigensolver (VQE), in which a quantum device prepares a parameterized ansatz state 
and measures expectation values, while a classical optimizer updates the circuit 
parameters~\cite{Peruzzo2014, Grimsley2019, Huggins2020, Tilly2022, Otten2022, Fedorov2022, ZhangAsthana2025}. 
Related hybrid quantum--classical optimization ideas have also been explored beyond electronic-structure settings, for 
example in combinatorial optimization problems~\cite{shaydulin2024evidence}.

More recently, a family of methods based on Rayleigh–Ritz subspace diagonalization has attracted increasing attention. 
The non-orthogonal variational quantum eigensolver (NOVQE) extends VQE by preparing multiple independently parameterized 
quantum ansatz states, measuring their Hamiltonian and overlap matrix elements, and classically solving a generalized 
eigenvalue problem~\cite{Huggins2020}. The optimization-free nonorthogonal quantum eigensolver (NOQE) removes the 
variational parameter-optimization loop by preparing quantum states with fixed parameters and directly solving the 
generalized eigenvalue problem~\cite{Baek2023NOQE}. Other related subspace methods—including quantum subspace expansion (QSE), 
quantum Krylov approaches, quantum filter diagonalization, and symmetry-structured variants such as Q-SENSE—differ primarily in how 
the quantum subspace is constructed, yet share a common hybrid workflow: quantum circuits prepare the basis states and estimate 
matrix elements, while classical routines perform the subsequent eigenvalue computation~\cite{McClean2017QSE, Parrish2019, Stair2019, Klymko2022, Yoshioka2025, cortes2022quantum, QSENSE2025}. 
In contrast, sample-based quantum diagonalization (SQD) adopts a different strategy: a quantum device generates a determinant 
subspace via sampling, while classical high-performance computing resources perform configuration recovery, subspace construction, 
and large-scale eigenvalue solving~\cite{RobledoMoreno2025,epperly2022theory,kanno2023quantum}.

Against this landscape, we introduce CANOE (Classically Assisted Non-Orthogonal Eigensolver), a distinct level of classical–quantum 
hybridization. Rather than generating the subspace entirely on quantum hardware or constructing it purely from sampled determinants, 
CANOE hybridizes at the level of basis construction itself: quantum-prepared states form the expressive core of the subspace, while a 
large set of classically constructed states provides the addition of variational freedom. As a consequence, a significant portion of 
the projected matrix elements can be evaluated efficiently on classical hardware, reducing the quantum sampling burden while 
retaining quantum-generated expressivity.

This architecture, the combination of many inexpensive classically constructed basis states with a smaller set of quantum-prepared 
states, necessitates calculating the overlap matrix elements between classical basis states and quantum-prepared states. A naive 
approach would reconstruct the quantum state via full state tomography and compute overlaps classically. However, full quantum state 
tomography requires resources that scale exponentially with the number of qubits, $O(4^n)$, rendering it infeasible beyond small 
systems~\cite{Haah2017}. This exponential scaling has motivated the development of more measurement-efficient techniques, such as 
classical shadow tomography, which enable estimation of many properties of a quantum state without reconstructing the full density 
matrix~\cite{Huang2020, Huang2021, Gresch2025, Ren2025}. To address this challenge, we develop a histogram-based sampling strategy 
that directly estimates the required matrix elements between classical determinants and quantum-prepared states, significantly 
reducing sampling overhead compared to tomography-based and shadow approaches.

In addition to measurement considerations, constructing a hybrid classical–quantum subspace introduces a numerical challenge. Quantum-
prepared states often exhibit substantial mutual overlap, and when combined with a large number of classical basis states, this can 
induce significant linear dependence within the total subspace. Such redundancy leads to ill-conditioning of the overlap matrix in 
the generalized eigenvalue problem, potentially amplifying sampling noise and causing numerical instability. To mitigate this issue, 
we adopt a Schur-complement-based formulation that avoids direct inversion of the ill-conditioned overlap matrix, thereby stabilizing 
the eigenvalue computation while preserving the physical content of the hybrid subspace~\cite{Golub2013, Saad2011, Knyazev2001}.

\begin{figure}[t]
    \centering
    \includegraphics[width=\linewidth]{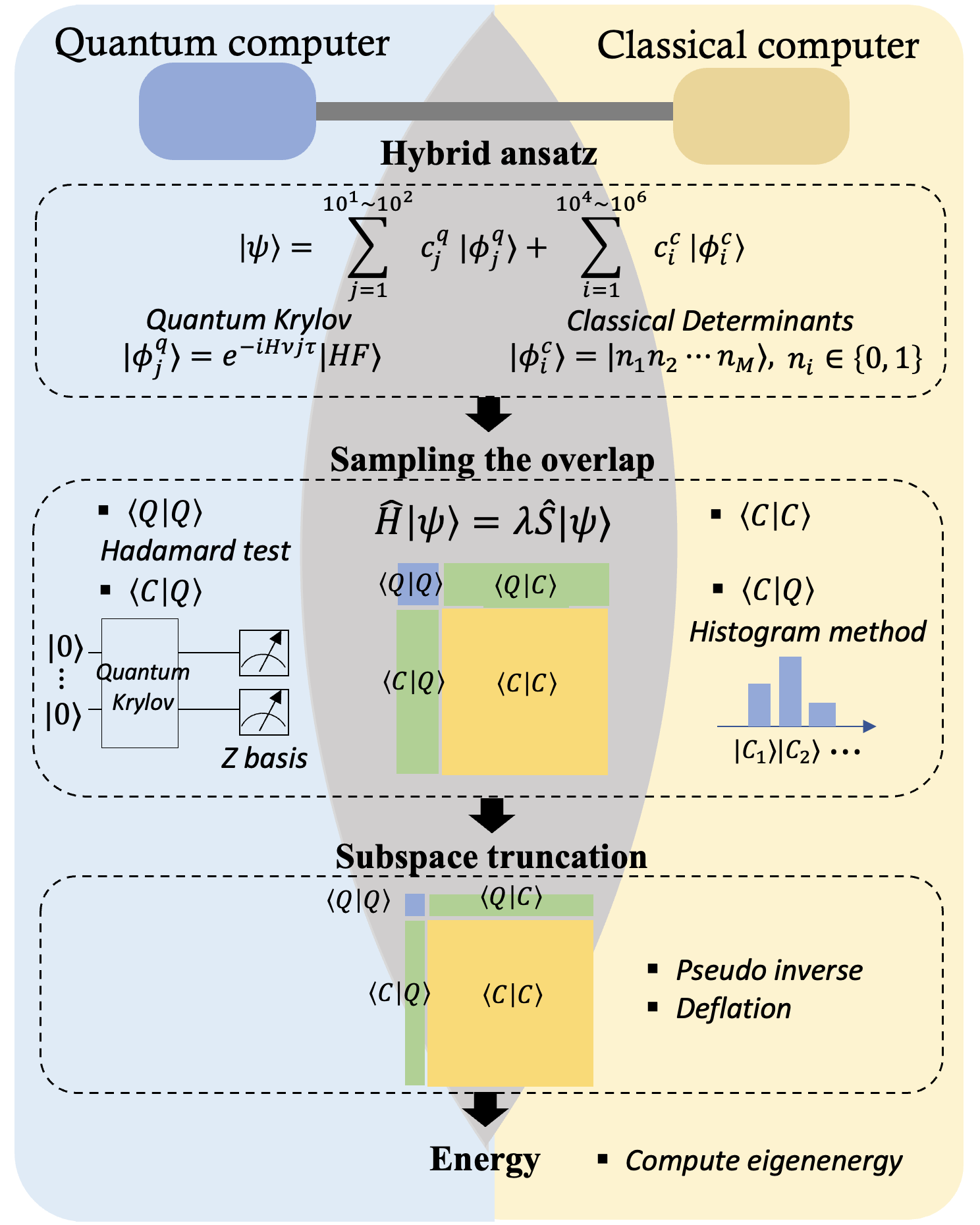}
    \caption{Overview of the CANOE protocol. The workload is distributed between classical (left) and quantum (right) processors. 
    \textbf{Hybrid ansatz:} CANOE combines a large number of classical determinants with a small number of quantum-prepared states. 
    \textbf{Sampling the overlap:} The projected matrices decompose into $\langle Q|Q\rangle$ (evaluated on quantum hardware using Hadamard test), $\langle C|C\rangle$ (computed on classical hardware), and $\langle Q|C\rangle$ blocks; the cross terms are estimated by sampling the quantum states and postprocessing classically via a histogram-based estimator. 
    \textbf{Subspace truncation:} Ill-conditioning of the overlap matrix is mitigated entirely on classical hardware using a Schur-complement formulation with deflation or pseudo-inverse stabilization.}
  \label{fig:schematic}
\end{figure}

In this work, we introduce the CANOE hybrid subspace framework and demonstrate its effectiveness for ground-state energy estimation 
across a range of molecular systems. The protocol is summarized in \cref{fig:schematic}. In \cref{sec:results}, we first analyze 
classical-quantum subspace hybridization and its representational capabilities through large-scale simulations of a strongly 
correlated 76-spin-orbital chromium benchmark system in an idealized infinite-shot limit. We then study sampling-based overlap 
estimation, focusing on a histogram-based method and on how sampling noise propagates to the ground-state energy error, and finally 
examine stabilization of the resulting generalized eigensolver for the non-orthogonal hybrid basis. In \cref{sec:discussion}, we 
discuss the main bottlenecks of CANOE and outline directions for further improvement. Detailed technical implementations of the 
hybrid ansatz construction, overlap-sampling protocol, and subspace-truncation strategy are presented in \cref{sec:methods}.

\section{Results}
\label{sec:results}

The objective of CANOE is to compute the ground-state energy of a second-quantized fermionic Hamiltonian to chemical accuracy. The 
electronic Hamiltonian is given by
\begin{equation}
    \hat{H} = h_q^p \hat{a}_p^\dagger \hat{a}_q 
    + \frac{1}{4} h_{qs}^{pr} \hat{a}_p^\dagger \hat{a}_r^\dagger \hat{a}_s \hat{a}_q ,
\end{equation}
where $h_q^p$ and $h_{qs}^{pr}$ denote the one- and antisymmetrized two-electron integrals, respectively. Repeated indices imply 
summation, and $\hat{a}_p^\dagger$ ($\hat{a}_p$) are fermionic creation (annihilation) operators acting on the corresponding spin 
orbitals. To solve the associated Schrödinger equation, CANOE constructs a hybrid classical–quantum subspace and performs a projected 
diagonalization within this basis.

\subsection{Classical-Quantum Subspace Hybridization}
\label{subsec:subhyb}

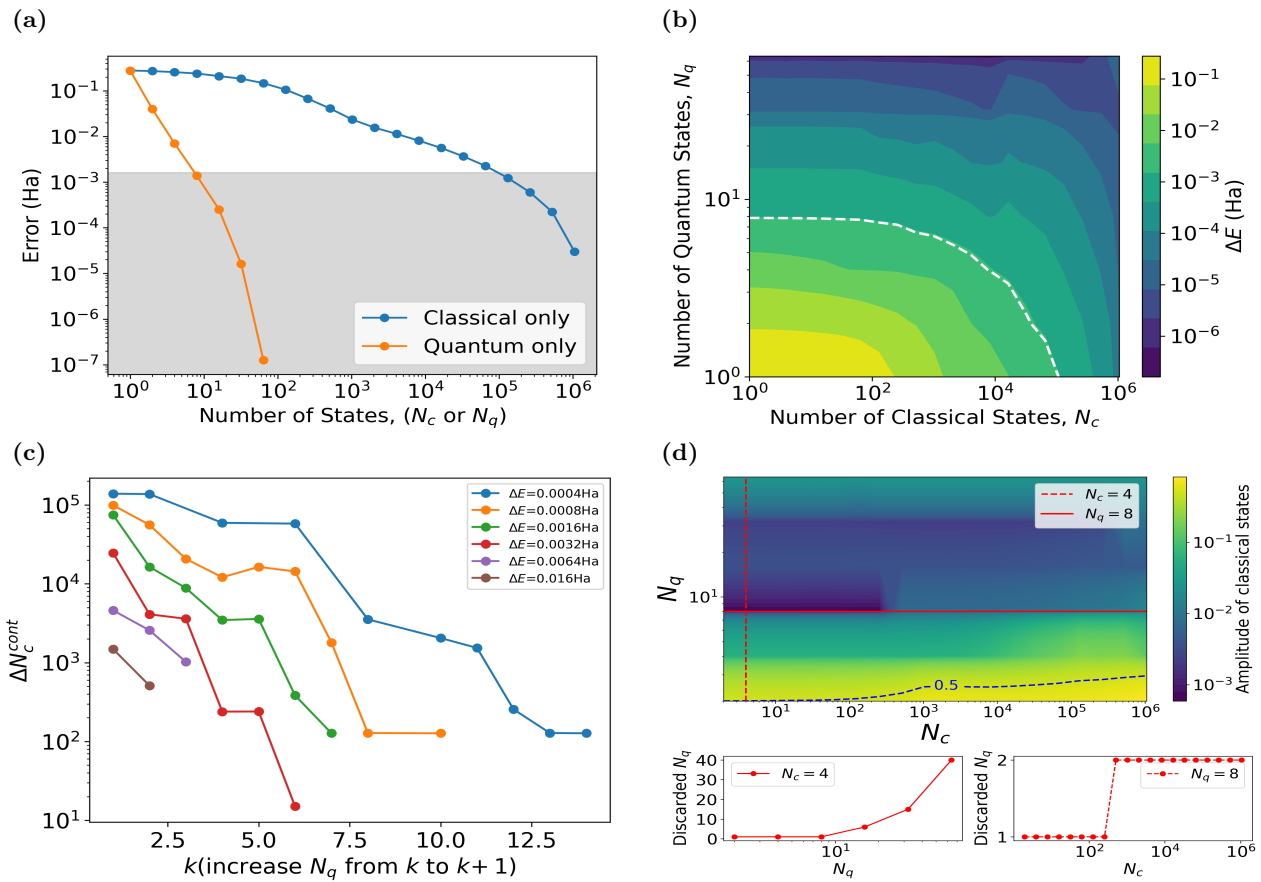
\begin{figure*}[t]
	\centering
		\begin{tikzpicture}
\begin{groupplot}[
    group style={
        group size=2 by 2,
        horizontal sep=10pt,
        vertical sep=5pt
    },
    width=0.55\linewidth,
    height=0.4\linewidth,
    axis lines=none,   
    ticks=none,
    enlargelimits=false,
    clip=false
]

\nextgroupplot
\addplot graphics[xmin=0, xmax=1, ymin=0, ymax=1]
{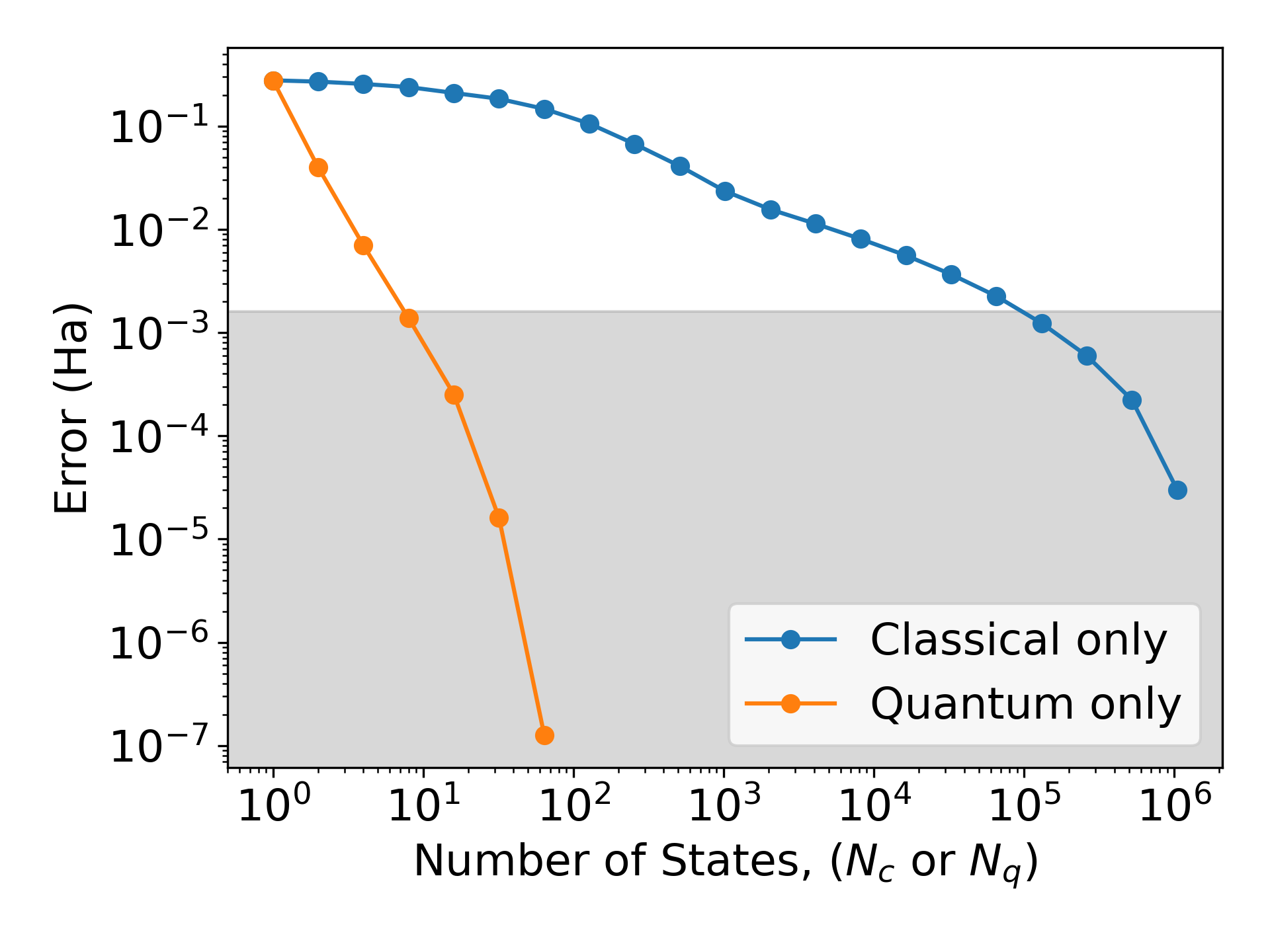};
\node[anchor=south west,font=\bfseries\small,fill=white,inner sep=1pt]
at (rel axis cs:0.02,1.00) {(a)};

\nextgroupplot
\addplot graphics[xmin=0, xmax=1, ymin=0, ymax=1]
{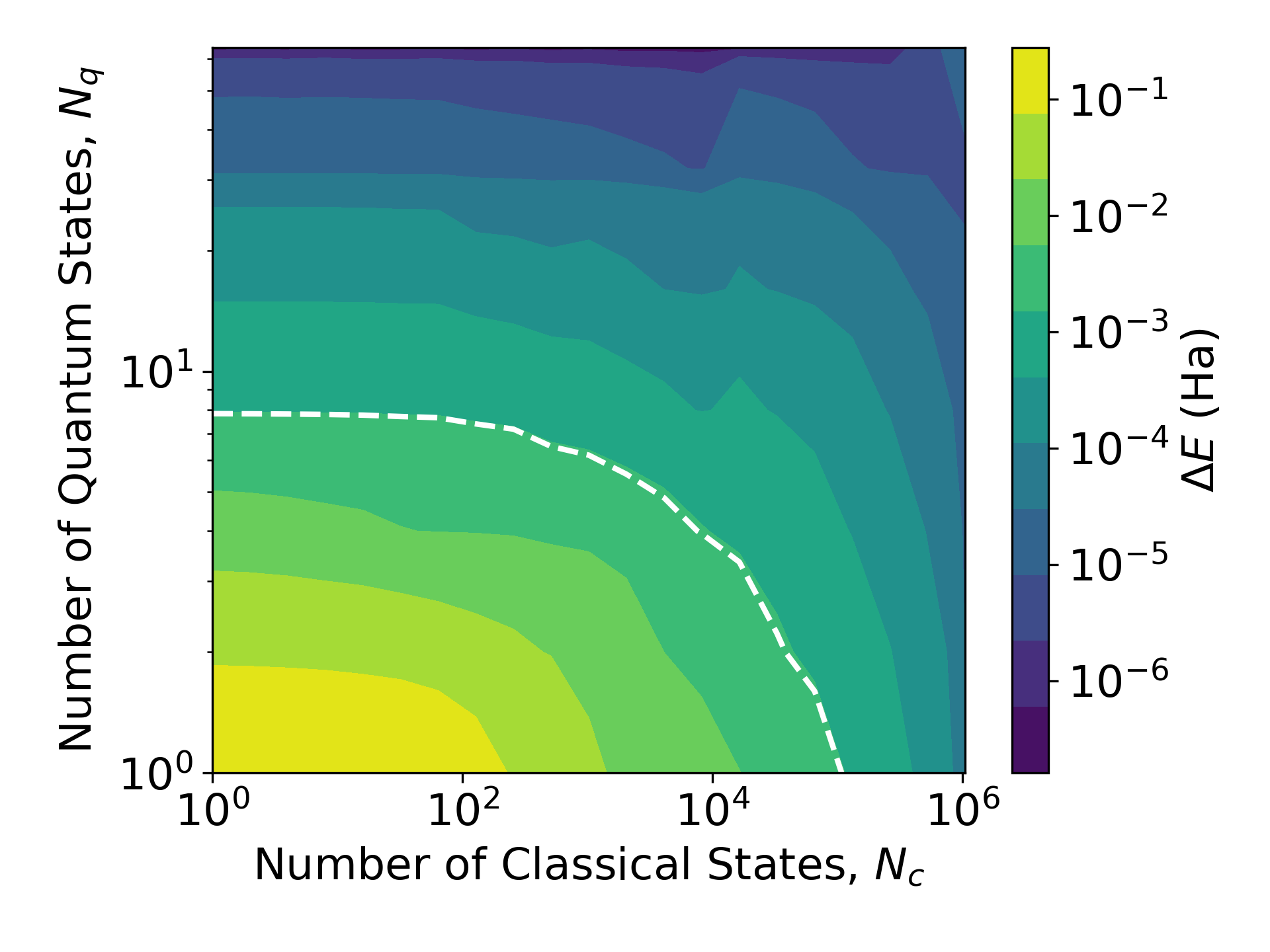};
\node[anchor=south west,font=\bfseries\small,fill=white,inner sep=1pt]
at (rel axis cs:0.02,1.00) {(b)};

\nextgroupplot
\addplot graphics[xmin=0, xmax=1, ymin=0, ymax=1]
{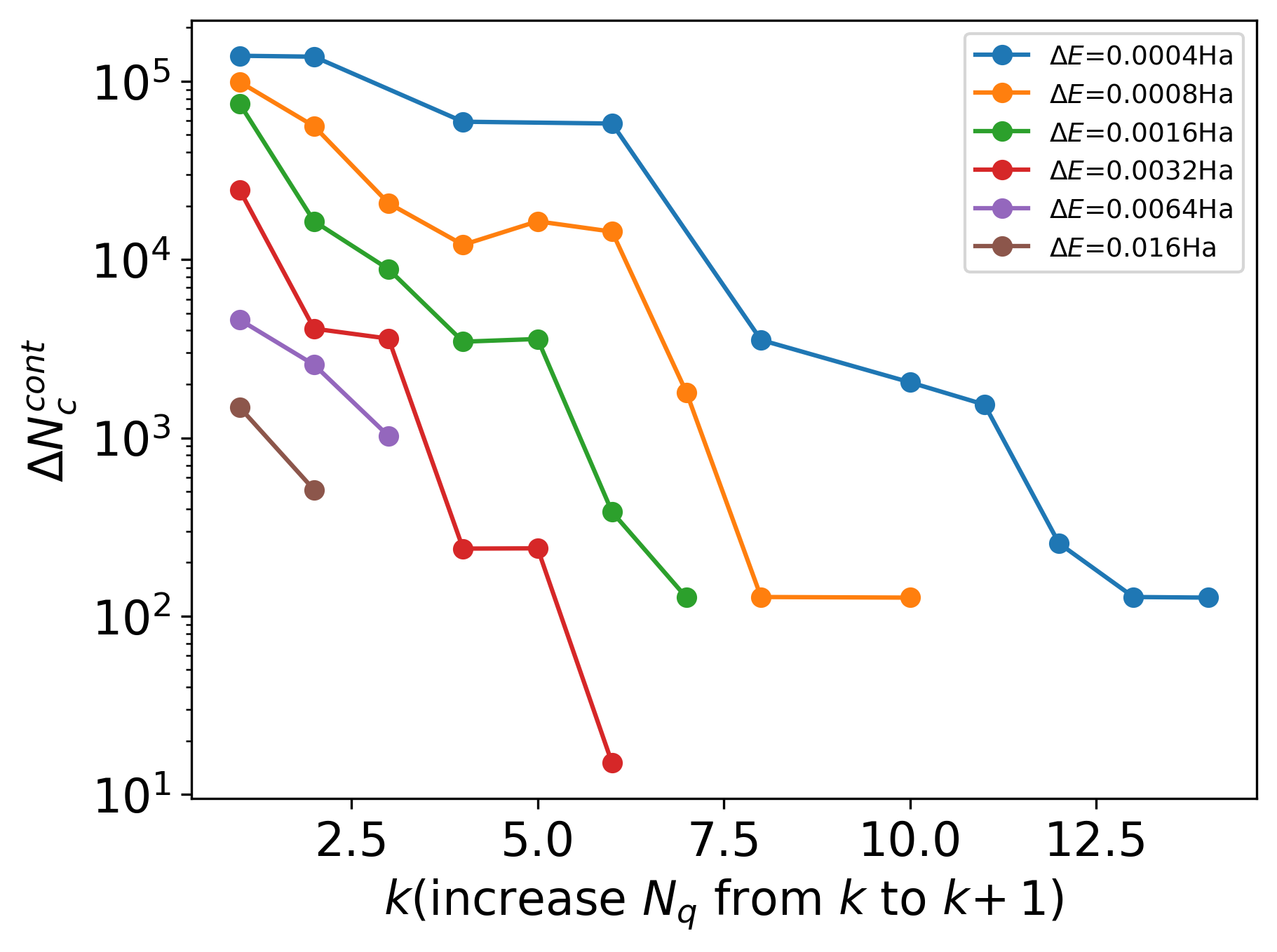};
\node[anchor=south west,font=\bfseries\small,fill=white,inner sep=1pt]
at (rel axis cs:0.02,1.00) {(c)};

\nextgroupplot
\addplot graphics[xmin=0, xmax=1, ymin=0, ymax=1]
{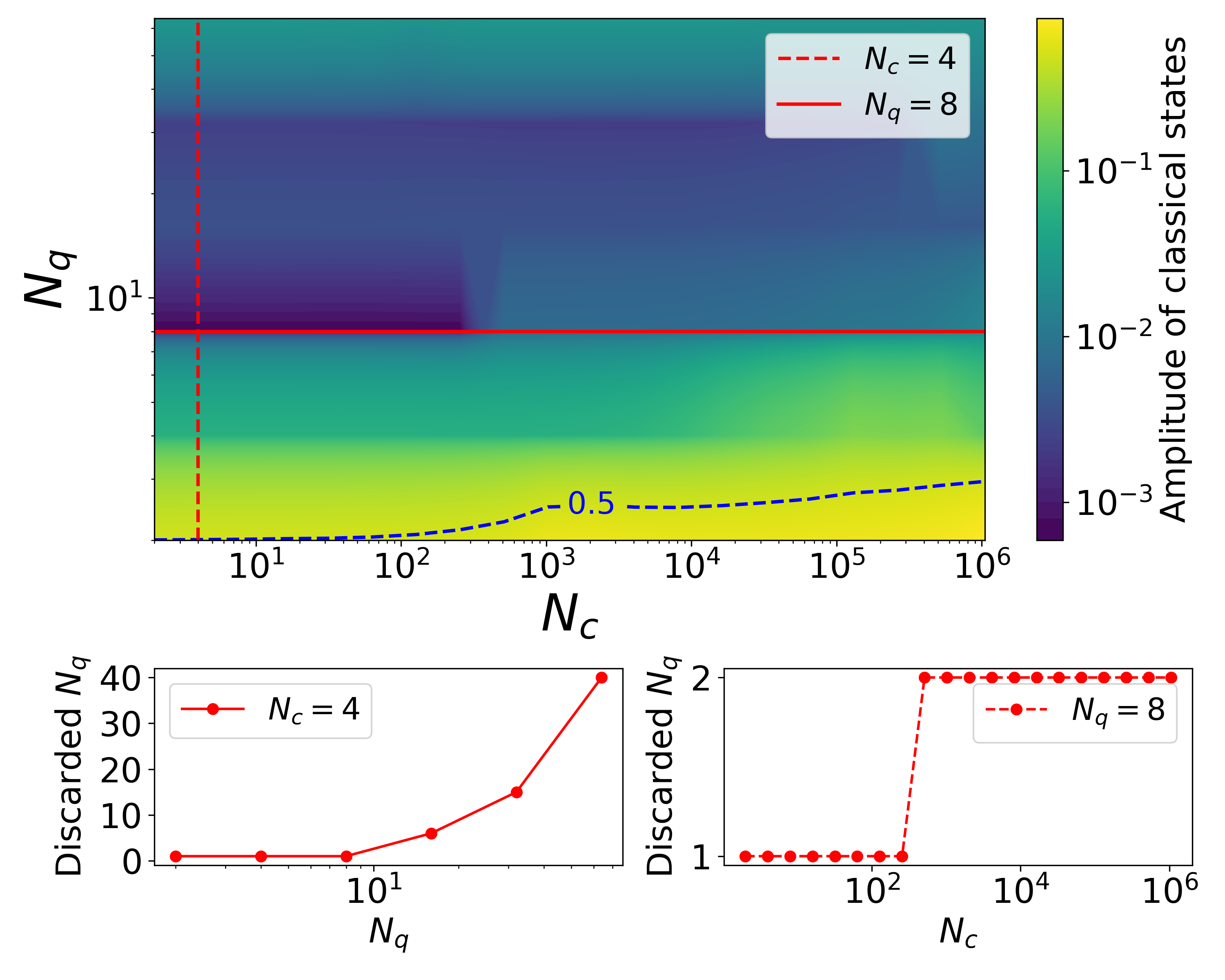};
\node[anchor=south west,font=\bfseries\small,fill=white,inner sep=1pt]
at (rel axis cs:0.02,1.00) {(d)};

\end{groupplot}
\end{tikzpicture}
		\caption{%
	    Classical-Quantum Subspace hybridization analysis on Cr system with 38 spatial orbitals and 14 electrons in cc-pVDZ-DK basis.
	    \textbf{(a)}: ground-state energy error (Ha) versus basis-set size for a purely classical SHCI determinant (blue) and a purely quantum Krylov subspace (orange). The gray band indicates ``chemical accuracy'' ($1.5936\times10^{-3}\,\mathrm{Ha}$).
	    \textbf{(b)}: ground-state energy error as a function of classical ($N_c$) and quantum ($N_q$) basis sizes. The white dashed line indicates the iso-error contour corresponding to chemical accuracy.
	    \textbf{(c)}: marginal reduction in the number of classical basis states, $\Delta N^{\text{cont}}_{\text{c}}$, that can be eliminated when one additional quantum basis state is introduced ($k\!\to\!k+1$) while remaining on the same ground-state energy-error contour.
	    \textbf{(d)}: ground-state classical-amplitude landscape over classical ($N_c$) and quantum ($N_q$) basis sizes. The color map shows the total classical amplitude $\sum_{i=1}^{N_c} |c_i|^{2}$ in the normalized ground state obtained from the generalized eigensolver. The contour labeled $0.5$ marks the boundary where the classical and quantum sectors contribute equal weight. The bottom left and bottom right insets show the number of discarded quantum states along the lines $N_c=4$ and $N_q=8$, respectively, in the upper panel.
}
	\label{fig:benchmark}	
\end{figure*}

The CANOE ansatz is defined as
\begin{equation}
|\psi\rangle
=
\sum_{i=1}^{N_c} c_i^{c} |\phi_i^c\rangle
+
\sum_{j=1}^{N_q} c_j^{q} |\phi_j^q\rangle ,
\end{equation}
where $|\phi_i^c\rangle$ and $|\phi_j^q\rangle$ denote classical and quantum basis states, respectively. 
\emph{Classical states} correspond to families of states for which the required overlap and 
Hamiltonian matrix elements remain classically tractable. In this work we restrict attention to determinants,
which, as we demonstrate, are especially efficient.
More generally, the classical sector could in principle include other states, such as Clifford states~\cite{ravi2022cafqa} 
or matrix product states~\cite{orus2014practical}.
\emph{Quantum states} are prepared and measured on a quantum processor. Quantum states can also be any states that are 
substantially more efficient to prepare on quantum hardware than to represent classically. For example, they may be generated by 
ans\"atze such as unitary coupled cluster singles doubles (UCCSD) or hardware-efficient 
circuits~\cite{Fedorov2022,Tilly2022,Kandala2017}. 
In practice, CANOE targets a large number ($10^4$–$10^6$) of classical basis states and a small number ($10$–$10^2$) of 
quantum basis states, reflecting realistic resource constraints of both classical and quantum hardware.
In the occupation-number representation over $M$ spin orbitals, the corresponding classical basis states are
\begin{equation}
|\phi_i^c\rangle = |n_1\, n_2\, \dots\, n_M\rangle, 
\quad n_k \in \{0,1\},\; i = 0,1,\dots,N_c,
\end{equation}

The quantum basis states are defined as
\begin{equation}
|\phi_j^q\rangle = e^{-i H\, t_j} |HF\rangle, 
\qquad j = 0,1,\dots,N_q,
\end{equation}
where  $H$ is the Hamiltonian of the system, $t_j = j\,\tau$ defines discrete time steps of size $\tau$, and $|HF\rangle$ 
is the Hartree--Fock state. 
These states form a non-orthogonal Krylov subspace generated by time evolution of the Hartree–Fock reference state, chosen because 
Hamiltonian-generated Krylov directions encode the system dynamics and can capture correlations that are difficult to reproduce within 
the truncated classical determinant space~\cite{Stair2019,Klymko2022,Shen2023RealTimeKrylov,cortes2022quantum}. The case $j=0$ 
corresponds to the Hartree–Fock state $|HF\rangle$, which is also contained within the classical determinant space.

Projecting the Schrödinger equation for $\hat{H}$ onto the hybrid basis $\{\ket{\phi_\mu}\} = \{\ket{\phi_i^c}\} \cup \{\ket{\phi_j^q}\}$ 
yields the generalized eigen-problem
\begin{equation}
\mathcal{H} \, \mathbf{c} = \lambda \, \mathcal{S} \, \mathbf{c},
\end{equation}
where $\mathcal{H}_{\mu\nu}=\bra{\phi_\mu}\hat{H}\ket{\phi_\nu}$ and $\mathcal{S}_{\mu\nu}=\braket{\phi_\mu|\phi_\nu}$.
The full workflow is summarized in \cref{fig:schematic}. 

To validate the CANOE protocol, we consider a strongly correlated chromium atom in the cc-pVDZ-DK basis 
with 38 spatial orbitals (76 spin orbitals) and 14 electrons, corresponding to a 76-qubit representation.
For computational efficiency, we only simulate the system in a restricted basis generated by the selection
of only the most important determinants via a selected heat-bath configuration interaction (SHCI)~\cite{li2018fast}
procedure, resulting in a sparse wavefunction simulator~\cite{mullinax2025large}.

Before analyzing hybridization, we first examine purely classical and purely quantum subspaces. 
\Cref{fig:benchmark}(a) shows the ground-state energy error as a function of basis size for each 
case. The quantum Krylov subspace converges significantly faster: slightly more than 8 quantum states 
are sufficient to reach chemical accuracy, whereas nearly $10^{5}$ classical determinants are required 
to achieve comparable precision. This highlights the strong expressive power of the quantum-generated directions.

We next vary both $N_c$ and $N_q$ simultaneously. \Cref{fig:benchmark}(b) presents the infinite-shot 
ground-state error over the $(N_c, N_q)$ plane. Consistent with \cref{fig:benchmark}(a), the error gradient is much steeper 
along the $N_q$ direction, indicating that adding quantum states yields a larger marginal improvement. Nevertheless, 
increasing either $N_c$ or $N_q$ improves the energy across the surface, demonstrating that classical and 
quantum basis states augment the subspace in complementary ways. The white dashed line marks the iso-error contour 
corresponding to chemical accuracy, thereby delineating the combinations of $(N_c,N_q)$ that are sufficient 
to reach this threshold. Around the elbow of this contour, near $N_q\sim 5$ and $N_c\sim 10^4$, the plot 
provides a useful example of the local tradeoff between the two sectors: as quantified more directly in 
 \cref{fig:benchmark}(c), increasing the quantum basis from $N_q=5$ to $N_q=6$ allows the classical basis 
size required for chemical accuracy to be reduced by roughly $3\times10^3$ determinants, whereas decreasing 
from $N_q=5$ to $N_q=4$ requires a comparable increase in $N_c$ to remain on the same contour. Thus, the elbow 
reflects a real change in the marginal replacement power of additional quantum states.

To quantify this complementarity, \cref{fig:benchmark}(c) analyzes the reduction in the number of classical 
determinants that can be eliminated when one additional quantum state is introduced while remaining on the same 
iso-error contour. The first added quantum state replaces the largest number of classical determinants, 
reflecting that it introduces a direction most orthogonal to the classical subspace. Subsequent quantum states 
exhibit diminishing replacement power, as the most significant missing correlations have already been captured. 
Notably, the replacement effect becomes more pronounced in the high-accuracy regime, where classical improvements 
alone become increasingly costly. At the same time, this increased replaceability can signal that newly added 
quantum directions are only weakly independent of an already enriched hybrid basis, leading to near-redundancy and 
motivating deflation of such directions in the generalized eigensolver, which we discuss in more depth in \cref{sec:gev}.

Finally, \cref{fig:benchmark}(d) examines how the ground-state amplitude is distributed between 
classical and quantum sectors. The color map shows the total classical weight $\sum_{i=1}^{N_c}|c_i|^2$ 
in the normalized ground state. For very small $N_q$, the classical sector carries most of the weight. 
However, as additional quantum states are introduced, the quantum sector rapidly accounts for a dominant 
portion of the expressivity. The first quantum state is omitted from the graph because it is identical to the 
Hartree--Fock reference already included in the classical determinant space. The non-smooth features observed 
in the landscape reflect truncation of nearly linearly dependent quantum directions. \Cref{sec:gev}
provides a detailed description of how we handle linear dependencies in the generalized eigenvalue problem,
which includes discarding linearly-dependent directions. Along fixed lines 
$N_c=4$ and $N_q=8$, we observe that the number of discarded quantum states increases as more are added, 
indicating growing redundancy. Therefore, increasing $N_q$ beyond a certain point does not necessarily 
yield proportional gains and must be balanced against resource cost. This also
points to the need for more efficient selection of quantum states, perhaps through choosing different time schedules.

\subsection{Overlap Estimation}
\label{subsec:oveest}
%

Since CANOE uses a hybrid classical-quantum subspace, the overlap matrix $\mathcal{S}$ naturally decomposes into three blocks: $\mathcal{S}^{cc}$, $\mathcal{S}^{cq}$, and $\mathcal{S}^{qq}$:

\begin{equation}
\mathcal{S}=
\begin{pmatrix}
\mathcal{S}^{cc} & \mathcal{S}^{cq} \\
\mathcal{S}^{qc} & \mathcal{S}^{qq},
\end{pmatrix},
\end{equation}
where
\begin{align}
\mathcal{S}^{cc}_{ik} = \braket{\phi_i^c | \phi_k^c}, \quad
\mathcal{S}^{cq}_{il} = \braket{\phi_i^c | \phi_l^q}, \\
\mathcal{S}^{qc}_{jk} = \braket{\phi_j^q | \phi_k^c}, \quad
\mathcal{S}^{qq}_{jl} = \braket{\phi_j^q | \phi_l^q}.
\end{align}

The $N_c^2$ classical--classical block is simply the identity matrix, and the $N_q^2$ quantum--quantum block can be 
measured using the Hadamard test in the quantum node, as is proposed in the standard NOQE scheme~\cite{Baek2023NOQE}. 
Our main focus is therefore the $N_c \times N_q$ classical--quantum block, namely how to efficiently estimate 
the projection of a quantum state onto a classical state, including both its magnitude and phase, without resorting 
to expensive quantum tomography or using a large number of Hadamard tests. 

As we assume a large number of classical states, it is desirable that the projection onto classical states can be 
obtained through classical post-processing, which significantly reduces the workload of the quantum computer. One 
recent approach is the classical shadow method~\cite{Huang2020, Ren2025}. However, local Pauli classical shadows rely 
on measurements in randomly chosen $X$, $Y$, and $Z$ bases, whereas in our problem we only need projections onto 
the $Z$ basis, since the determinants are defined in that basis. Furthermore, as the sample complexity of classical 
shadow shows, the required number of samples scales with the shadow norm, which grows exponentially with the number of 
qubits for the specific operators we seek to estimate. The dependence on the number of target observables is only 
logarithmic; in our setting, because the target observables are the determinant projections associated with the 
classical sector, this contribution scales as $\log N_c$. A full analysis of the sample complexity for Pauli-based 
classical shadows is derived in \cref{appendix:classical_shadow}. For larger systems the total sampling cost
required to estimate all of these determinant projections can still become significant. Therefore, we propose a 
method optimized specifically for projections onto the $Z$-basis space, which we 
call \emph{histogram-based overlap estimation}.

\subsubsection{Histogram-based overlap estimation method}

There are many possible approaches to overlap estimation; for example, recent work has considered 
time-series-based estimation of overlaps such as the Loschmidt amplitude~\cite{Patel2026QuantumPhaselift}.

To calculate the overlap, let

\begin{equation}
\alpha_s:=\braket{s|\phi_j^q},\qquad \beta_s:=\braket{s|\chi},
\end{equation}
\noindent where $\ket{s}$ represents a determinant and $\ket{\phi_j^q}$ a quantum basis state. Here, $\ket{\chi}$ is a batch consisting of the equal superposition of the classical states:
$\ket{\chi_k}=\frac{1}{\sqrt{m}}\sum_{s\in S_k}\ket{s}$, where
$m:=|S_k|$ is the batch size and $k$ indexes the equal-size batches. Thus,
the amplitudes $\beta_s$ are known and the number of batches is
$B=\lceil N_c/m \rceil$. Define the interference states as
\begin{equation}
\ket{\psi_R}:=\frac{1}{\sqrt{2}}\left(\ket{\phi_j^q}+\ket{\chi}\right),\qquad
\ket{\psi_I}:=\frac{1}{\sqrt{2}}\left(\ket{\phi_j^q}+i\,\ket{\chi}\right),
\label{eq:psi_ri_def}
\end{equation}
and denote the computational-basis histograms by

\begin{align}
p_q(s):=&\Pr(s|\phi_j^q)=|\alpha_s|^2,\\
p_R(s):=&|\langle s | \psi_R\rangle|^2,\\
p_I(s):=&|\langle s | \psi_I \rangle|^2.
\end{align}

Here, note that $p_R(s)$ and $p_I(s)$ are not probabilities because $\ket{\psi_R}$ and $\ket{\psi_I}$ 
are not normalized states. Therefore, to obtain these quantities, we should renormalize the 
probabilities, as shown in \cref{sec:methods}. Then the per-string unbiased estimator for $\alpha_s$ is

\begin{equation}
\begin{aligned}
\hat{\alpha}_s
=& \frac{
\left[p_R(s)-\frac{1}{2}\left(p_q(s)+|\beta_s|^2\right)\right]
}{\bar{\beta}_s} \\
&+ i\,\frac{
\left[p_I(s)-\frac{1}{2}\left(p_q(s)+|\beta_s|^2\right)\right]
}{\bar{\beta}_s}.
\end{aligned}
\end{equation}
where $\bar{\beta}_s$ denotes the complex conjugate of $\beta_s$, i.e., $\bar{\beta}_s=\beta_s^*$.

In general, the batch size $m$ introduces a trade-off: larger $m$ reduces the number of batches 
$B=\lceil N_c/m \rceil$ and hence the number of processing steps, but it also decreases $|\beta_s|=1/\sqrt{m}$, 
making the estimator more sensitive to statistical noise and potentially increasing the estimation error.
In our histogram-method simulations, we chose $m=5000$ as a practical compromise, since it keeps $B$ small 
enough to reduce runtime while maintaining $|\beta_s|=1/\sqrt{5000}\approx 1.4\times 10^{-2}$, which was 
empirically sufficient to retain chemical-accuracy-level performance.

\subsubsection{From overlap to Hamiltonian matrix element}

Using the property that the action of a Pauli string on a computational basis state yields another 
computational basis state with a new coefficient, we can obtain the Hamiltonian matrix element 
without additional quantum measurements, using only the elements of the overlap matrix. For a 
Hamiltonian decomposed as $H=\sum_{\gamma=1}^{n_H} h_{\gamma}P_{\gamma}$, we have

\begin{align}
\mathcal{H}_{ij}&=\braket{\phi_i^c|H|\phi_j^q}=\sum_{\gamma=1}^{n_H} h_{\gamma} \braket{\phi_i^c|P_{\gamma}|\phi_j^q}\\
&=\sum_{\gamma=1}^{n_H} h_{\gamma} \theta_{\gamma}(s_i)^{*} \braket{s_i\oplus b_{\gamma}|\phi_j^q},
\label{eq:H_ij}
\end{align}
where $s_i$ represents a bit string of $\ket{\phi_i^c}$, $\theta_{\gamma}(s_i)=i^{m_{\gamma}}(-1)^{z_{\gamma}\cdot s_i}$, $z_{\gamma}\cdot s_i=\sum_{k}z_{\gamma}(k)s_k$ $m_{\gamma}$ is the number of $Y$ gate in the Pauli string $P_{\gamma}$, and $P_{\gamma}$, $b_{\gamma}$, and $z_{\gamma}$ are defined as

\begin{align}
P_{\gamma} =& \bigotimes_{k=1}^{n_{\mathrm{qubit}}} \sigma_{\gamma k}, \quad \sigma_{\gamma, k}\in {I,X,Y,Z},\\
b_\gamma(k)=&\begin{cases}
1,\quad \sigma_{\gamma_k}\in\{X,Y\}\\
0,\quad \text{otherwise}
\end{cases},
\\
z_\gamma(k)=&\begin{cases}
1,\quad \sigma_{\gamma_k}\in\{Z,Y\}\\
0,\quad \text{otherwise}
\end{cases}.
\end{align}

Since $\braket{s_i\oplus b_{\gamma}|\phi_j^q}$ is an overlap between a quantum basis state and a determinant, 
we can extract this value from the overlap matrix, or simply set it to zero if the corresponding overlap 
element is absent. This method allows us to use measurements of the quantum state in the 
computational basis, which may already be available from measurements of the quantum--quantum block, to infer 
the values of all classical states and Hamiltonian matrix elements which have support on the quantum states.

\begin{figure*}[t]
	\centering
	\begin{tikzpicture}
\begin{groupplot}[
    group style={
        group size=2 by 2,
        horizontal sep=10pt,
        vertical sep=0pt
    },
    width=0.55\linewidth,
    height=0.4\linewidth,
    axis lines=none,
    ticks=none,
    enlargelimits=false,
    clip=false
]

\nextgroupplot
\addplot graphics[xmin=0, xmax=1, ymin=0, ymax=1]
{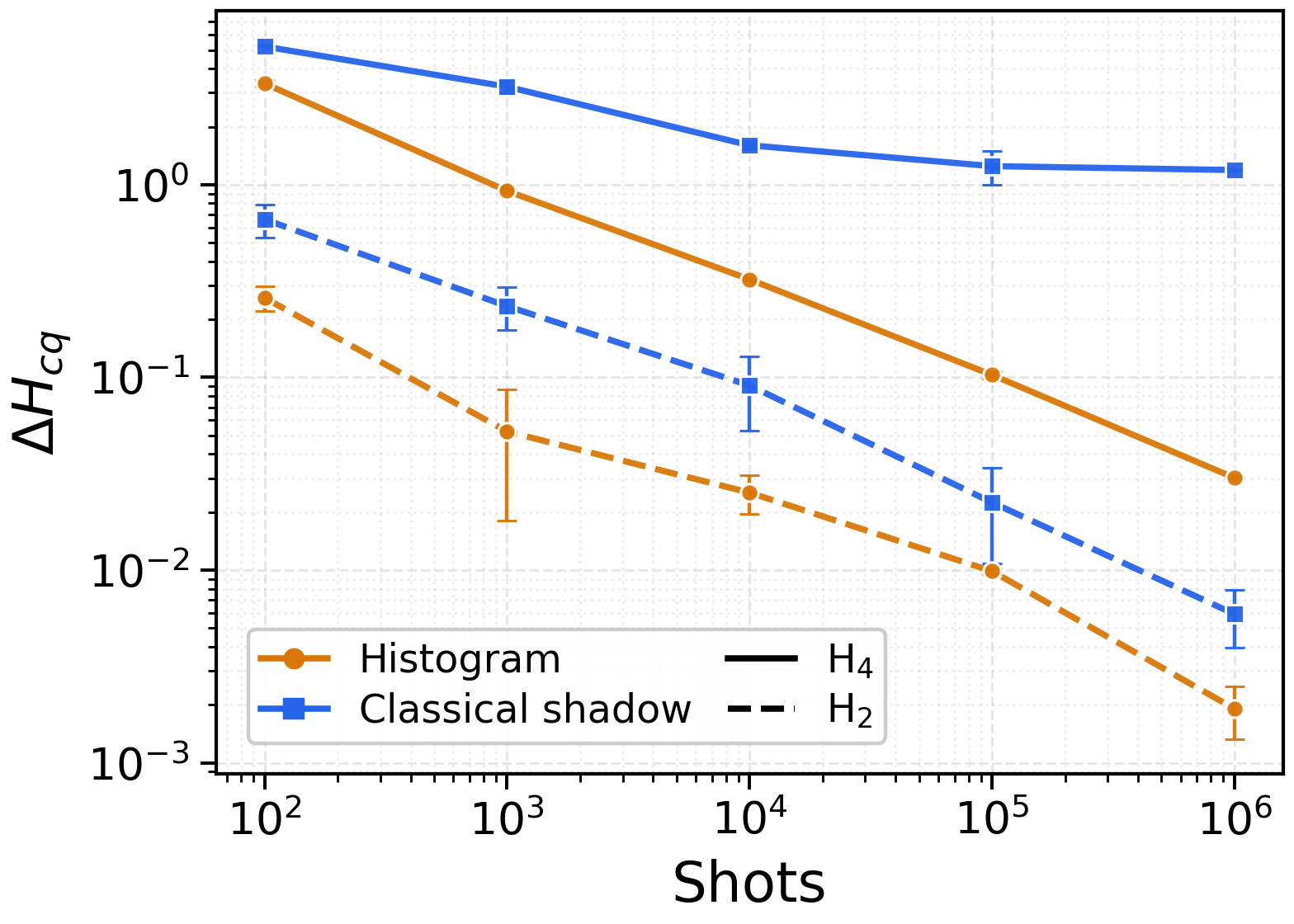};
\node[anchor=south west,font=\bfseries\small,fill=white,inner sep=1pt]
at (rel axis cs:0.02,1.00) {(a)};

\nextgroupplot
\addplot graphics[xmin=0, xmax=1, ymin=0, ymax=1]
{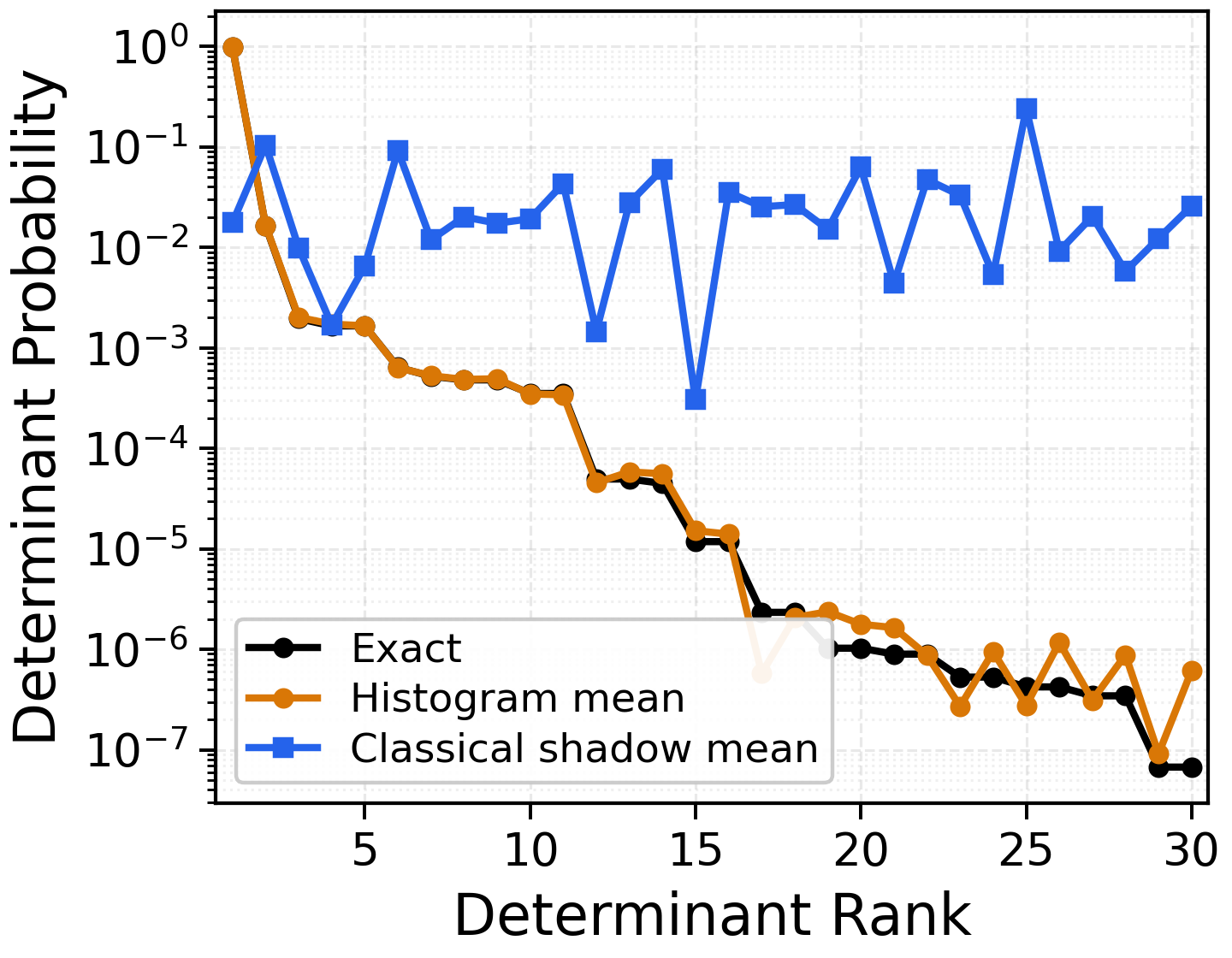};
\node[anchor=south west,font=\bfseries\small,fill=white,inner sep=1pt]
at (rel axis cs:0.02,1.00) {(b)};

\nextgroupplot
\addplot graphics[xmin=0, xmax=1, ymin=0, ymax=1]
{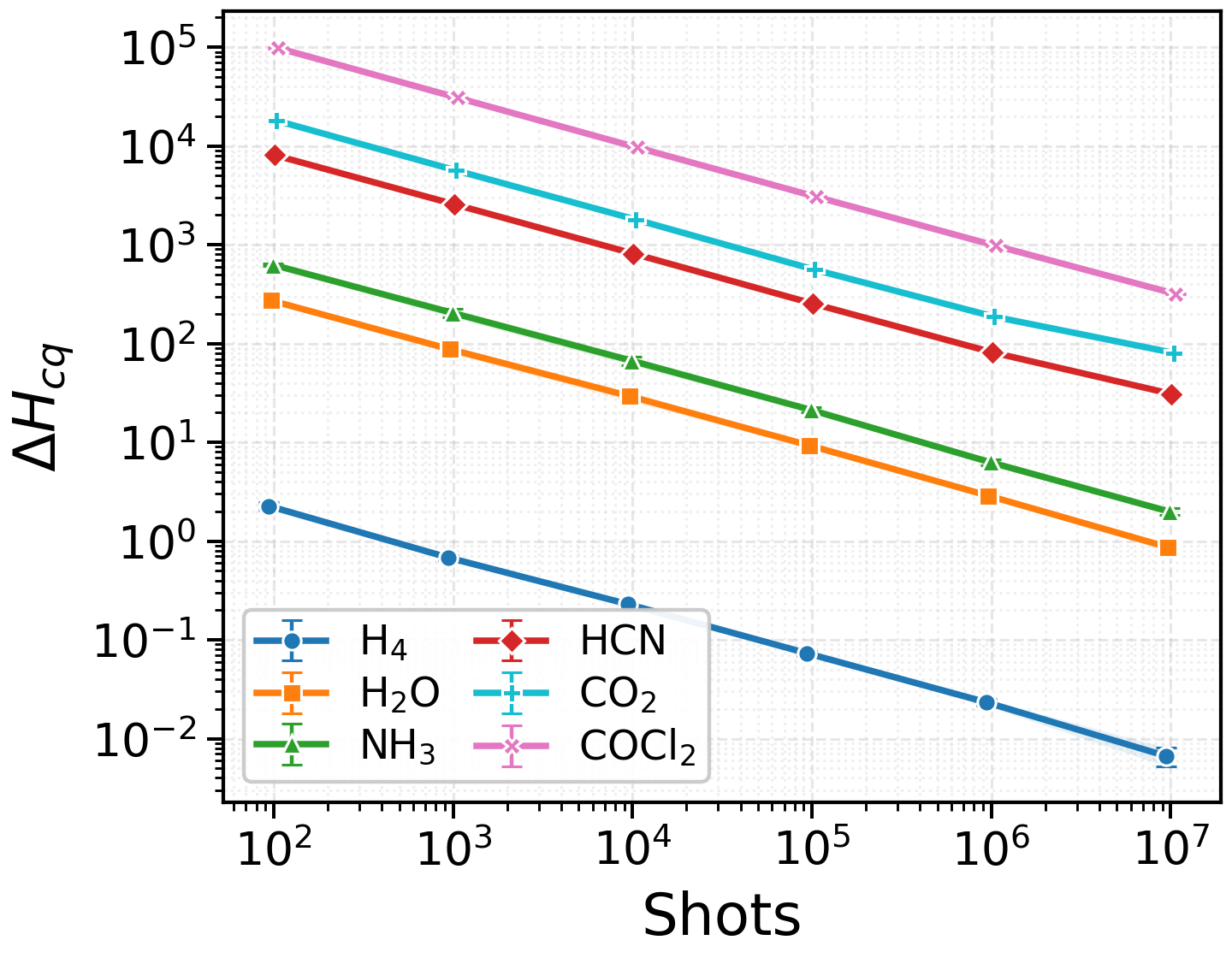};
\node[anchor=south west,font=\bfseries\small,fill=white,inner sep=1pt]
at (rel axis cs:0.02,1.00) {(c)};

\nextgroupplot
\addplot graphics[xmin=0, xmax=1, ymin=0, ymax=1]
{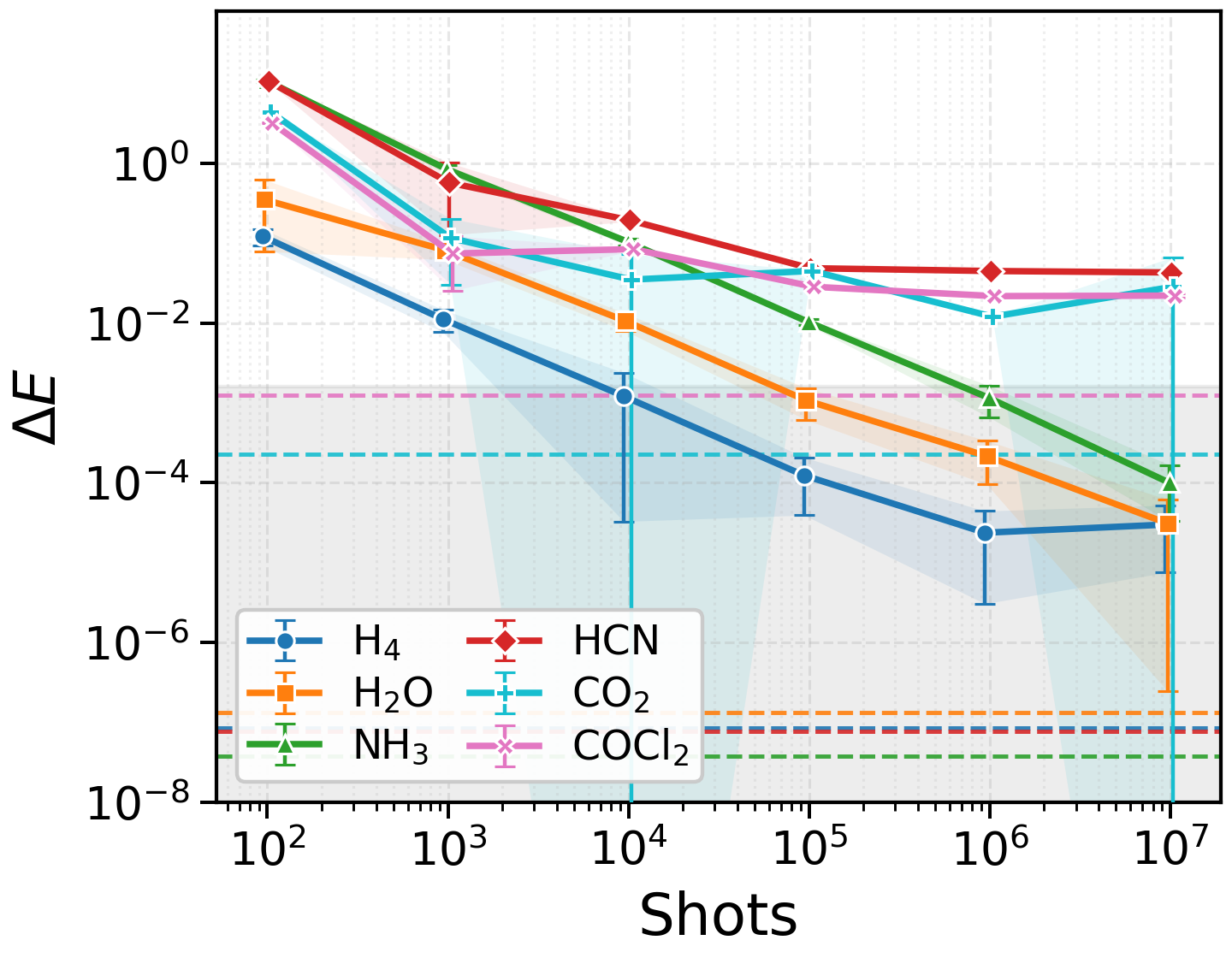};
\node[anchor=south west,font=\bfseries\small,fill=white,inner sep=1pt]
at (rel axis cs:0.02,1.00) {(d)};

\end{groupplot}
\end{tikzpicture}
    \caption{%
    Numerical benchmark of histogram-based overlap estimation.
    \textbf{(a)}: Comparison between histogram-based and classical-shadow estimation for the Hamiltonian-matrix error $\Delta H$ as a function of the number of shots for H$_4$ (solid) and H$_2$ (dashed). The reference values are taken from the infinite-shot limit of the sampling simulation. For H$_2$, we use $N_c=2$(the full SHCI space, used here to test sampling noise) and $N_q=2$, while H$_4$ uses the same basis sizes as in Table~II.
    \textbf{(b)}: Determinant probabilities versus determinant rank for $\mathrm{H}_4$, shown at $10^6$ shots, comparing the histogram-based estimation and classical shadow against the ``exact'' result, where ``exact'' denotes the infinite-shot limit.
    \textbf{(c)}: Classical--quantum Hamiltonian-matrix error $\Delta H_{cq}$ obtained using the histogram-based estimation method as a function of the number of shots for the benchmark systems. The reference values are obtained from analytically computed Hamiltonian-matrix elements, corresponding to the infinite-shot limit.
    \textbf{(d)}: Ground-state energy error $\Delta E$ as a function of the number of shots for the same benchmark systems. The error is defined relative to the reference ground state energy, which is SHCI variational energy. The dashed lines indicate the gap between the SHCI variational energy and the infinite-shot-limit CANOE energy in the truncated subspace. The infinite-shot-limit ground-state energy is obtained by loading the exact truncated $(H,S)$ matrices, Hermitizing them, solving $Hc=ESc$ with LOBPCG, deflation, and pseudo-inverse over a threshold sweep, and comparing the resulting solver outputs with $E_{\mathrm{true}}$.
}
	\label{fig:overlap_benchmark}	
\end{figure*}
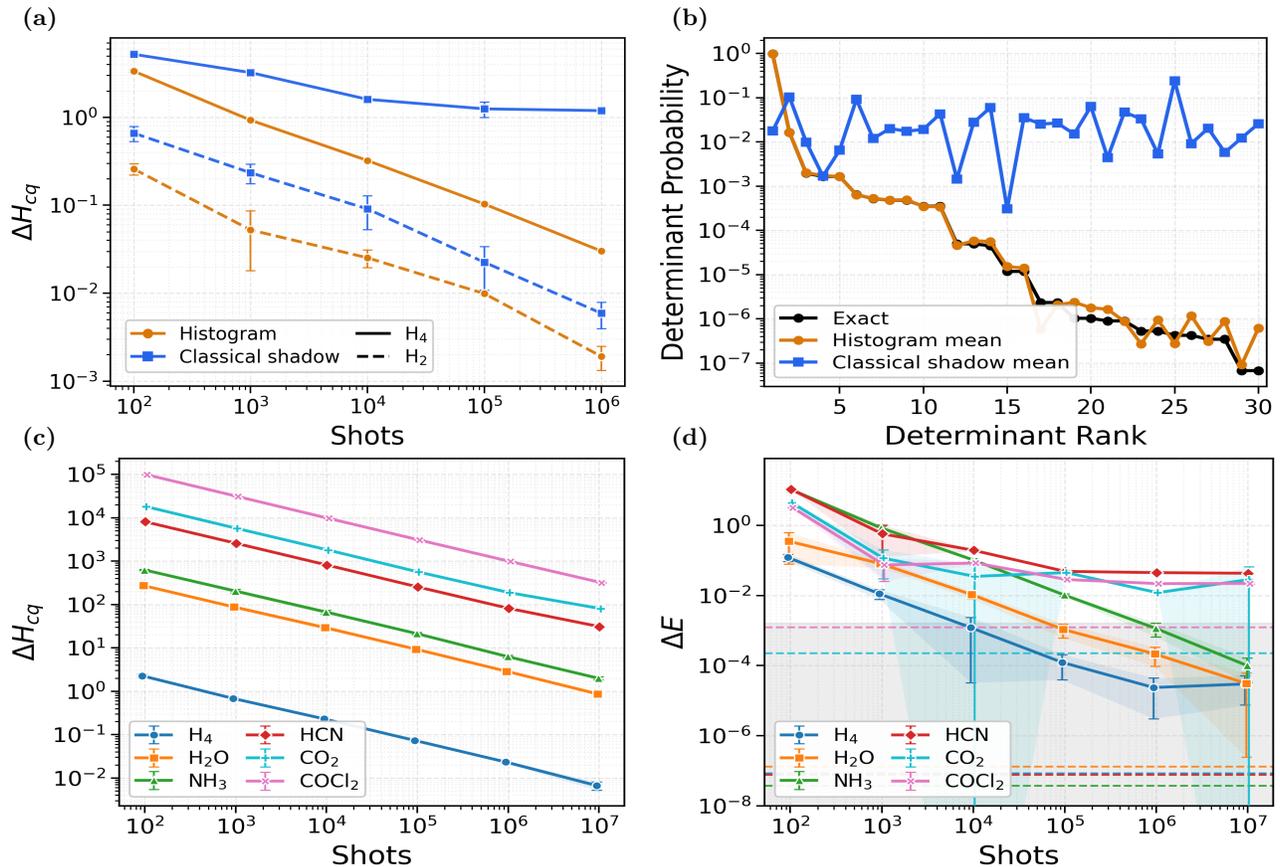

\subsubsection{Numerical benchmark of histogram-based estimation}
\label{subsec:numerics}
To assess the practical performance of histogram-based overlap estimation, we numerically simulate 
sampled overlap matrices, post-process them into Hamiltonian matrices, and solve the resulting 
generalized eigenvalue problem with the locally optimal block preconditioned conjugate-gradient (LOBPCG) eigensolver~\cite{Knyazev2001,Saad2011,Golub2013}. Before turning to the larger benchmark 
set, we first compare histogram-based estimation with classical shadows 
on the $\mathrm{H}_4$ system. Although $\mathrm{H}_4$ is small, it already provides a clear illustration 
of the different noise sensitivities of the two methods. \cref{fig:overlap_benchmark}(a)
shows how the error of the reconstructed Hamiltonian matrix changes with the number of shots.

A direct comparison of raw shot counts requires some care, because a ``shot'' has a different meaning 
for the two methods. For classical shadows, one shot corresponds to a measurement of a Pauli operator 
for a given quantum state, whereas for histogram-based estimation one shot corresponds to a sample 
contributing to the computational-basis histogram of a given quantum state. For the full classical--quantum 
block, the total histogram shot count is therefore the per-histogram shot count multiplied by $N_q(1+2B)$, 
because each quantum basis state requires one reference histogram and $2B$ interference histograms. 
A fair comparison should therefore be based on the total shot complexity required to achieve a fixed error 
$\epsilon$ in an individual classical--quantum Hamiltonian matrix element $\mathcal{H}_{ij}$, as summarized in 
\cref{tab:shot_complexity}. The full derivation of these sample complexities can be found in 
\cref{appendix:sample_complexity}. Both methods avoid an explicit linear dependence on $N_c$, the number of 
classical states, which is desirable in the CANOE regime where $N_c$ is intentionally large. 
The key difference lies in the dependence on system size: classical shadows carry an exponential 
factor in the number of spin orbitals (equivalently qubits), whereas histogram-based estimation 
scales only linearly through the batching factor. This scaling difference explains the gap observed in 
\cref{fig:overlap_benchmark}(a) and is consistent with the much milder behavior of classical shadows in 
smaller systems such as $\mathrm{H}_2$.

\begin{table}[t]
\centering
\normalsize
\setlength{\tabcolsep}{3pt}
\renewcommand{\arraystretch}{1.05}
\resizebox{\columnwidth}{!}{%
\begin{tabular}{@{}cc@{}}
\toprule
\textbf{Method} & \textbf{Total shots} \\
\midrule
Hadamard test ~\cite{Baek2023NOQE} &
$\mathrm{O}\!\left(\frac{N_c N_q n_H\lVert\mathbf{H}\rVert_{2}^2}{\epsilon^2}\right)$ \\
\midrule
Classical shadow ~\cite{Huang2020, Ren2025} &
$\mathrm{O}\!\left(\frac{N_q\log(N_c)\lVert\mathbf{H}\rVert_{1}^2 3^{n_{\mathrm{qubit}}}}{\epsilon^2}\right)$ \\
\midrule
Histogram-based estimation &
$\mathrm{O}\!\left(\frac{N_q(1+2B) m \lVert\mathbf{H}\rVert_{1}^2}{\epsilon^2}
\log\!\left(2^{n_{\mathrm{qubit}}} N_q(1+2B)\right)\right)$ \\
\bottomrule
\end{tabular}}
\caption{Sample complexity of the overlap-estimation methods for constructing the full classical--quantum ($c$--$q$) block. Here, $N_c$ and $N_q$ denote the numbers of classical and quantum states, respectively. $m$ is the batch size. $B=\lceil N_c/m\rceil$ is the number of batches. $n_H$ is the number of Pauli terms in the Hamiltonian. $\lVert\mathbf{H}\rVert_{1}$ and $\lVert\mathbf{H}\rVert_{2}$ are the corresponding coefficient one-norm and two-norm. $\epsilon$ is the target error in an individual classical--quantum Hamiltonian matrix element $\mathcal{H}_{ij}$. $\delta$ is the failure probability.}
\label{tab:shot_complexity}
\end{table}

\cref{fig:overlap_benchmark}(b) shows the same distinction at the level of determinant probabilities 
for $\mathrm{H}_4$: histogram-based estimation reproduces the ground-state distribution more 
faithfully than the classical-shadow approach.

We then benchmark histogram-based estimation across the molecular set $\mathrm{H}_4$, 
$\mathrm{H}_2\mathrm{O}$, $\mathrm{NH}_3$, $\mathrm{HCN}$, $\mathrm{CO}_2$, and $\mathrm{COCl}_2$, 
using the simulation protocol described in \cref{sec:methods}. The benchmark systems are summarized 
in \cref{tab:benchmark_systems}. In \cref{fig:overlap_benchmark}(c), the error in the classical--quantum 
block of the Hamiltonian matrix decreases toward zero as the number of shots increases, approximately 
following the expected $1/\sqrt{N_{\mathrm{shot}}}$ scaling of the sampling model. The energy error in 
\cref{fig:overlap_benchmark}(d) is governed by three factors: sampling error, truncated-subspace gap, 
and instability of the generalized eigensolver. First, the sampling error decreases systematically as 
the number of shots increases, as seen in \cref{fig:overlap_benchmark}(c). Second, the dashed lines in 
\cref{fig:overlap_benchmark}(d) show the truncation gap, namely the difference between the 
reference energy $E_{\mathrm{true}}$ and the infinite-shot-limit CANOE energy within the truncated 
subspace; if the truncated space does not represent the ground state sufficiently well, increasing the
number of shots cannot overcome this floor. 
Third, the stability of the generalized eigensolver is particularly critical.

\begin{table}[t]
\centering
\normalsize
\setlength{\tabcolsep}{3pt}
\renewcommand{\arraystretch}{1.05}
\begin{tabular*}{\columnwidth}{@{\extracolsep{\fill}}ccccc@{}}
\toprule
\textbf{System} & \textbf{(so,e)} & \textbf{$N_c$} & \textbf{SHCI} & \textbf{$N_q$} \\
\midrule
$\mathrm{H_4}$ & (8,4) & 10 & 36 & 4 \\
$\mathrm{H_2O}$ & (14,10) & 100 & 133 & 10 \\
$\mathrm{NH_3}$ & (16,10) & 1000 & 1259 & 10 \\
$\mathrm{HCN}$ & (22,14) & 10000 & 11712 & 64 \\
$\mathrm{CO_2}$ & (30,22) & 10000 & 46167 & 64 \\
$\mathrm{COCl_2}$ & (56,48) & 10000 & 98437 & 64 \\
\bottomrule
\end{tabular*}
\caption{Benchmark molecules used to simulate histogram sampling. The notation $(\mathrm{so},e)$ denotes the numbers of spin orbitals and electrons, respectively. $N_c$ and $N_q$ are the numbers of classical and quantum states, respectively. SHCI means the number of determinants of SHCI variational space.}
\label{tab:benchmark_systems}
\end{table}

\cref{fig:overlap_benchmark}(d) shows that the ground-state energy error does not decrease 
uniformly across systems. The reason is that the energy is obtained from the generalized eigenproblem,
$E_0=\frac{\mathbf{c}_0^\dagger \mathcal{H}\mathbf{c}_0}{\mathbf{c}_0^\dagger \mathcal{S}\mathbf{c}_0}$, 
rather than from an individual matrix element, where $\mathbf{c}_0$ is the lowest generalized-eigenvalue 
vector. As a result, the same level of sampling noise in the matrix elements of $\mathcal{H}$ and $\mathcal{S}$ 
can produce very different energy errors for different systems. This sensitivity can become more pronounced 
when important components of $\mathbf{c}_0$ lie along small-eigenvalue directions of $\mathcal{S}$ that are 
not well resolved at the current shot count; in that regime, near-linear dependencies can make the generalized 
eigensolver less stable and lead to plateaus or delayed convergence toward the infinite-shot limit. 
Therefore, the shot count required to reach chemical accuracy depends not only on the sample-complexity 
scaling, but also on the spectral conditioning of the truncated subspace for the specific system and the 
particular qualities of the generalized eigensolver used.

In the present numerical benchmarks, we introduced sampling noise only in the classical--quantum block. 
In a full quantum implementation, however, the quantum--quantum block would also need to be evaluated. 
One direct option is the Hadamard test. Alternatively, if determinant amplitudes are reconstructed for 
each quantum basis state on a common determinant set and in a consistent phase convention, then both diagonal
and off-diagonal quantum--quantum overlap and Hamiltonian matrix elements can in principle be assembled 
classically from those amplitudes, as shown in \cref{appendix:qq_from_overlaps}. In practice, 
this reconstruction is exact only when the determinant set is sufficiently complete for the relevant 
Pauli-shifted support, so the Hadamard test remains a direct route when that condition is not met.

%







\subsection{Generalized eigensolver for non-orthogonal basis}\label{sec:gev}

Since CANOE includes quantum basis states, the resulting basis is generally non-orthogonal and 
may even contain near-linear dependencies. This creates a numerical difficulty for iterative 
solvers of generalized eigenvalue problems, such as LOBPCG or Jacobi--Davidson, because they 
require repeated applications of $\mathcal{S}^{-1}$ or an equivalent stable factorization of the overlap 
matrix~\cite{Knyazev2001,Saad2011,Golub2013}. When $\mathcal{S}$ is nearly singular, standard 
Cholesky or $\mathrm{LDL}^{\mathsf T}$ factorizations can become unstable~\cite{Saad2011,Golub2013}. 
To avoid both the ill-conditioning of $\mathcal{S}$ and explicit diagonalization of the full 
large-scale problem, we exploit the block structure induced by the classical--quantum partition 
and work with the Schur complement of the quantum sector. This leads to two practical stabilization 
strategies: deflation of nearly dependent quantum directions and a pseudo-inverse-based preconditioner for LOBPCG.

With the classical states ordered first and the quantum states second, the overlap matrix can be written as
\begin{equation}
\mathcal{S}
=
\begin{pmatrix}
I & U \\
U^\dagger & M
\end{pmatrix},
\end{equation}
where $I$ is the classical--classical overlap block of size $N_c\times N_c$, $U$ is the classical--quantum 
overlap block of size $N_c\times N_q$, and $M$ is the quantum--quantum overlap block of size $N_q\times N_q$. 
The corresponding Schur complement is
\begin{equation}
\mathcal{S}_{\mathrm{schur}} = M - U^\dagger U.
\end{equation}
We diagonalize this matrix as
\begin{equation}
\mathcal{S}_{\mathrm{schur}}
=
V \Lambda V^\dagger,
\qquad
\Lambda=\mathrm{diag}(\lambda_1,\ldots,\lambda_{N_q}).
\end{equation}

In the deflation approach, we discard eigenvectors whose eigenvalues satisfy $\lambda_i < \tau_{\mathrm{rank}}$ 
and solve the problem in the reduced quantum subspace. Because $N_q$ is typically small in our setting, 
explicit full diagonalization of $\mathcal{S}_{\mathrm{schur}}$ is a reasonable way to implement this 
deflation step. In the pseudo-inverse approach, we construct $\mathcal{S}_{\mathrm{schur}}^{+}$ by 
inverting only the retained eigenvalues and use the resulting operator to build the LOBPCG preconditioner; 
in this case, the weak directions are flattened rather than deleted. The hybrid workflow used in 
practice is summarized in \cref{alg:schur_hybrid_solver}.

\medskip
\refstepcounter{algorithm}\label{alg:schur_hybrid_solver}
\begingroup
\noindent\begin{minipage}{\columnwidth}
\hrule

\vspace{4pt}
\textbf{Algorithm \thealgorithm: Schur-complement hybrid solver}

\vspace{2pt}
\hrule

\vspace{4pt}
\footnotesize
\begin{algorithmic}[1]
\Require $H=\begin{psmallmatrix}H_{cc} & H_{cq}\\ H_{qc} & H_{qq}\end{psmallmatrix}$, \quad $S=\begin{psmallmatrix}I & U\\ U^{\dagger} & M\end{psmallmatrix}$
\Require stabilization mode $\in \{\textbf{\emph{Pseudo-inverse}},\textbf{\emph{Deflation}}\}$, rank threshold $\tau_{\mathrm{rank}}$, initial block $X_0$
\State $S_{\mathrm{schur}} \gets M - U^{\dagger}U$
\State Diagonalize $S_{\mathrm{schur}} = V\,\mathrm{diag}(\lambda_1,\ldots,\lambda_{N_q})\,V^{\dagger}$
\If{mode = \textbf{\emph{Pseudo-inverse}}}
    \State $\mu_i \gets \begin{cases}
    1/\lambda_i, & \lambda_i > \tau_{\mathrm{rank}},\\
    0, & \lambda_i \le \tau_{\mathrm{rank}}
    \end{cases}$
    \State $S_{\mathrm{schur}}^{+} \gets V\,\mathrm{diag}(\mu_1,\ldots,\mu_{N_q})\,V^{\dagger}$
    \State \texttt{P\_inv} $\approx\begin{psmallmatrix}
    I + U S_{\mathrm{schur}}^{+} U^{\dagger} & -U S_{\mathrm{schur}}^{+}\\
    -S_{\mathrm{schur}}^{+} U^{\dagger} & S_{\mathrm{schur}}^{+}
    \end{psmallmatrix}$
    \State \texttt{A\_use} $\gets H$, \quad \texttt{B\_use} $\gets S$, \quad \texttt{X\_use} $\gets X_0$
\EndIf
\If{mode = \textbf{\emph{Deflation}}}
    \State $\mathcal{I}_{+} \gets \{\, i \mid \lambda_i > \tau_{\mathrm{rank}} \,\}$, \quad $V_{+} \gets V(:,\mathcal{I}_{+})$
    \State $U' \gets U V_{+}$, \quad $M' \gets V_{+}^{\dagger} M V_{+}$
    \State $H'_{cq} \gets H_{cq}V_{+}$, \quad $H'_{qc} \gets V_{+}^{\dagger}H_{qc}$, \quad $H'_{qq} \gets V_{+}^{\dagger}H_{qq}V_{+}$
    \State Build the reduced pair
    \Statex \hspace{\algorithmicindent} \texttt{A\_use} $\gets H'=\begin{psmallmatrix}H_{cc} & H'_{cq}\\ H'_{qc} & H'_{qq}\end{psmallmatrix}$
    \Statex \hspace{\algorithmicindent} \texttt{B\_use} $\gets S'=\begin{psmallmatrix}I & U'\\ (U')^{\dagger} & M'\end{psmallmatrix}$
    \State \texttt{P\_inv} $\approx (S')^{-1}$
    \State If $X_0=\begin{psmallmatrix}X_{0,c}\\X_{0,q}\end{psmallmatrix}$, set \texttt{X\_use} $\gets \begin{psmallmatrix}X_{0,c}\\V_{+}^{\dagger}X_{0,q}\end{psmallmatrix}$
\EndIf
\Statex \hspace{\algorithmicindent} \texttt{lobpcg(A\_use, X\_use, B=B\_use, M=P\_inv, ...)}
\end{algorithmic}

\vspace{4pt}
\hrule
\end{minipage}
\endgroup

For the results in \cref{subsec:subhyb}, we used the hybrid solver summarized above. By contrast, for the results in
\cref{subsec:oveest}, we used LOBPCG without additional deflation or pseudo-inverse regularization. 
In that setting, the sampled overlap matrices are perturbed by statistical noise away from the exact, 
highly linearly-dependent structure, which partially alleviates the ill-conditioning of $\mathcal{S}$,
as shown in \cref{tab:sinv_lambdamin}. 
In the sampled-noise regime considered here, truncating near-singular directions can distort the ground-state 
solution more than it improves the conditioning of the generalized eigenvalue problem. The appropriate 
treatment of the generalized eigenvalue problem should therefore be adapted to the noise level of the 
overlap matrix rather than fixed uniformly across all regimes.

\begin{table}[t]
\centering
\small
\setlength{\tabcolsep}{2.5pt}
\renewcommand{\arraystretch}{1.05}
\begin{tabular*}{\columnwidth}{@{\extracolsep{\fill}}ccc@{}}
\toprule
\textbf{Molecule} & \textbf{Analytic $1/|\lambda_{\min}|$} & \textbf{Sampled $1/|\lambda_{\min}|$} \\
\midrule
$\mathrm{H_4}$ & $\infty$ & $539.00\,(\pm 264.03)$ \\
$\mathrm{H_2O}$ & $\infty$ & $150.84\,(\pm 21.71)$ \\
$\mathrm{NH_3}$ & $\infty$ & $47.39\,(\pm 8.96)$ \\
$\mathrm{HCN}$ & $7.48 \times 10^{12}$ & $8.33\,(\pm 0.70)$ \\
$\mathrm{CO_2}$ & $5.82 \times 10^{12}$ & $2.69\,(\pm 0.06)$ \\
$\mathrm{COCl_2}$ & $5.31 \times 10^{12}$ & $7.49\,(\pm 0.21)$ \\
\bottomrule
\end{tabular*}
\par\smallskip
\refstepcounter{table}\label{tab:sinv_lambdamin}
\parbox{\columnwidth}{\raggedright\small\textbf{TABLE \thetable.} Comparison of the conditioning of analytic and sampled overlap matrices, measured by $1/|\lambda_{\min}|$. The sampled values denote the mean over the $10^7$-shot data.}
\end{table}

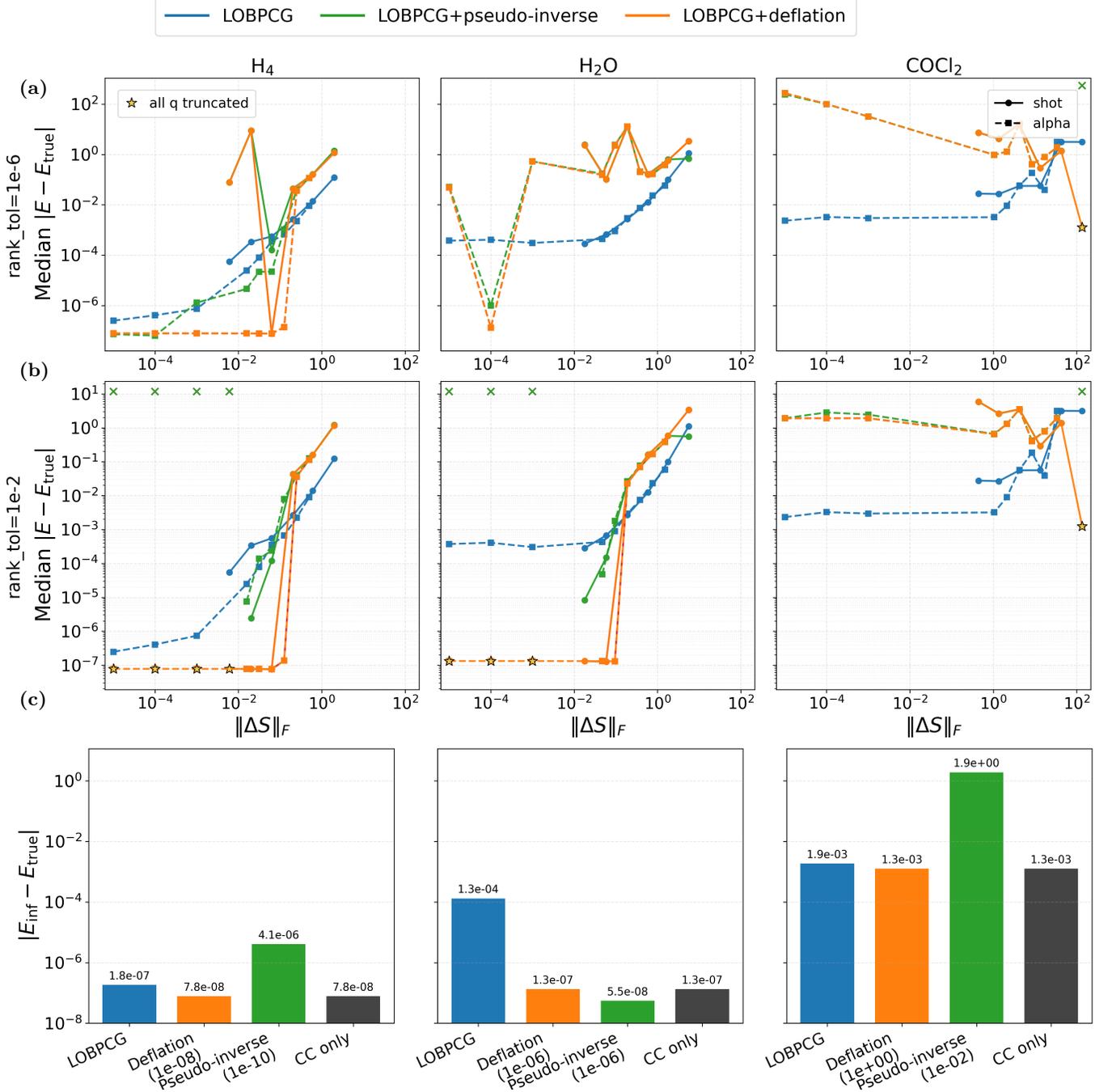
\begin{figure*}[t]
    \centering
    \begin{tikzpicture}
    \node[inner sep=0] (fig4) {\includegraphics[width=\textwidth]{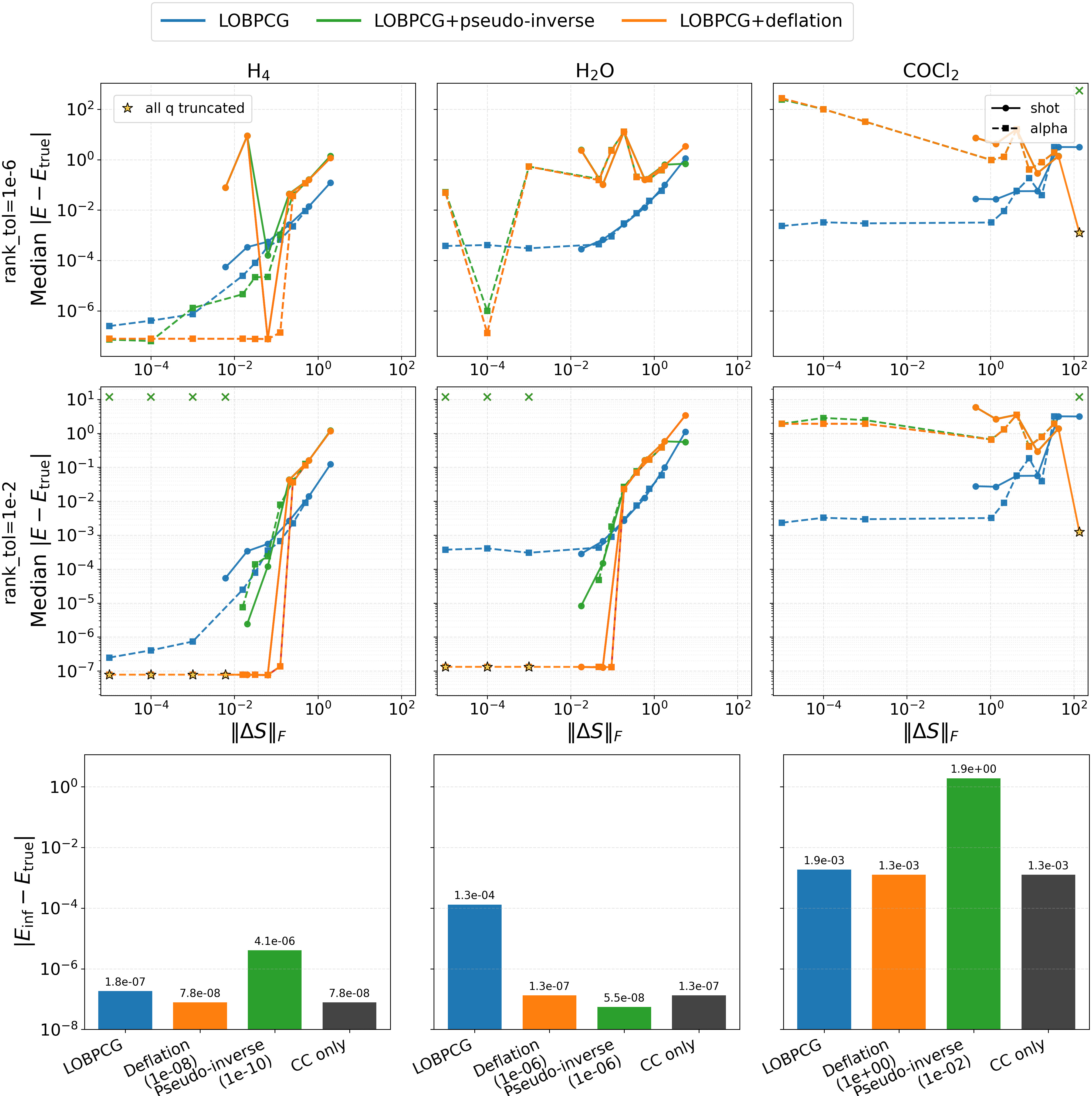}};
    \node[anchor=west,font=\bfseries\small,fill=white,inner sep=1pt] at ([xshift=6pt,yshift=-44pt]fig4.north west) {(a)};
    \node[anchor=west,font=\bfseries\small,fill=white,inner sep=1pt] at ([xshift=6pt,yshift=-176pt]fig4.north west) {(b)};
    \node[anchor=west,font=\bfseries\small,fill=white,inner sep=1pt] at ([xshift=6pt,yshift=-330pt]fig4.north west) {(c)};
    \end{tikzpicture}
    \caption{%
    Benchmark of eigensolver performance versus overlap-matrix noise for
    $\mathrm{H_4}$, $\mathrm{H_2O}$, and $\mathrm{COCl_2}$.
    Columns correspond to the three molecular systems.
    \textbf{(a)} and \textbf{(b)} show the median ground-state energy error,
    $|E - E_{\mathrm{true}}|$, as a function of $\|\Delta S\|_F$ for
    $\tau_{\mathrm{rank}} = 10^{-6}$ and $10^{-2}$, respectively, where
    $E$ denotes the ground-state energy estimated from the noisy matrices.
    In these two rows, solid curves denote the sampled-matrix benchmark and
    dashed curves denote the synthetic $\alpha$-noise benchmark. Blue, green,
    and orange denote LOBPCG, LOBPCG+pseudo-inverse, and LOBPCG+deflation,
    respectively. Green ``x'' markers indicate cases in which the pseudo-inverse
    solver failed to return a result, and star-shaped markers indicate cases in
    which all quantum directions were truncated so that only the classical
    subspace remained.
    \textbf{(c)} shows the corresponding infinite-shot-limit benchmarks,
    $E_{\mathrm{inf}} - E_{\mathrm{true}}$, together with the
    classical-classical only (CC-only) reference for each system shown in
    black.
    Values closer to zero indicate better agreement with the SHCI variational
    reference energy.}
    \label{fig:h2o_eigensolver_noise}
\end{figure*}

To make this dependence explicit, we parameterized the noise level by the Frobenius norm 
$\|\Delta S\|_{F}$ of the overlap-matrix perturbation. We varied this quantity in two ways: 
first, by associating each finite-shot sampled matrix with its corresponding $\|\Delta S\|_{F}$ value; 
and second, by constructing synthetic overlap matrices of the form 
$S(\alpha)=S+\alpha\bigl(S^{(N_0)}-S\bigr)$, where $S^{(N_0)}$ denotes a reference sampled matrix 
with $N_0=100000$, so that $\alpha$ continuously controls $\|\Delta S\|_{F}$. We then repeated the 
analysis across different Schur-complement thresholds and for three eigensolver variants: 
LOBPCG, LOBPCG with pseudo-inverse regularization, and LOBPCG with deflation. The resulting 
$\mathrm{H_4}$, $\mathrm{H_2O}$, and $\mathrm{COCl_2}$ benchmarks are shown in \cref{fig:h2o_eigensolver_noise}.

Several important aspects of the eigensolver behavior can be seen in \cref{fig:h2o_eigensolver_noise}. 
First, the eigensolvers respond differently to the Schur-complement threshold. In the sampled-matrix 
regime, the more aggressive threshold, $\tau_{\mathrm{rank}} =10^{-2}$, is more robust than 
$\tau_{\mathrm{rank}} = 10^{-6}$ over a broad range of noise levels. This indicates that, 
for sampled matrices, it is often advantageous to remove weakly independent or poorly resolved quantum 
directions more strictly. One likely reason is that finite-shot noise can render the sampled overlap 
matrix indefinite, so the generalized eigenproblem is no longer strictly variational. In this regime, 
adding more noisy quantum directions does not necessarily improve the ground-state estimate; instead, 
truncating unstable directions can yield a better result, and in the extreme case a classical-only 
solve can outperform a noisy $(c+q)$ calculation.

Second, the figure shows that LOBPCG behaves continuously and stably across the sampled-noise regime. 
Among the tested methods, it often provides the most consistent performance as the noise level varies, 
and in many sampled cases it gives the smallest error. This is consistent with the observation that 
finite-shot noise shifts the smallest overlap modes away from zero in magnitude, thereby relieving 
the severe near-singularity of the analytic overlap matrix; see \cref{tab:sinv_lambdamin}, where the 
sampled matrices exhibit substantially better conditioning with respect to the smallest eigenvalue. 
As a consequence, LOBPCG can remain numerically stable on sampled matrices even though the corresponding 
infinite-shot-limit generalized eigenproblem is substantially harder.

Third, the infinite-shot-limit benchmark in \cref{fig:h2o_eigensolver_noise}(c) shows 
that the quantum subspace can improve 
the ground-state energy when it is treated stably. For $\mathrm{H_2O}$, the pseudo-inverse solver gives 
a smaller gap in the infinite-shot limit than the classical-only reference, demonstrating that the 
quantum directions contain useful variational information. However, this improvement is not fully 
recovered in the sampled case at the same threshold, showing that noise can mask the benefit of the 
quantum subspace.

Finally, the infinite-shot-limit benchmarks of $\mathrm{H_4}$ and $\mathrm{COCl_2}$ show that, 
even in the infinite-shot limit, where the variational principle should hold, the classical-only solution 
can still be the most accurate among the tested eigensolver outputs. This should not be interpreted 
as evidence that the quantum subspace is intrinsically unhelpful. Rather, it indicates that the current 
eigensolver cannot always resolve weakly independent quantum directions reliably in a nearly singular 
generalized eigenproblem.

The overall picture therefore involves two distinct mechanisms: in the sampled regime, 
noise can destroy the variational structure and make truncation beneficial; in the infinite-shot-limit regime, 
the main limitation is the numerical resolution of weakly independent quantum directions.

%
\section{Discussion}
\label{sec:discussion}
%
We introduced CANOE as a hybrid non-orthogonal eigensolver that combines a
small number of quantum-prepared states with a much larger set of classical
determinants. In the ideal hybrid-basis setting, these two sectors play
complementary roles: the quantum states contribute variational directions that
are difficult to reproduce classically, while the classical sector enlarges the
representation space around them at comparatively low cost. To make this hybrid
basis realizable, we introduced histogram-based overlap estimation for
determinant-space ($Z$-basis) projections and used Schur-complement truncation
to stabilize the resulting generalized eigenvalue problem. Within this
framework, our numerical results reach chemical accuracy for the smaller
benchmark systems $\mathrm{H}_4$, $\mathrm{H}_2\mathrm{O}$, and $\mathrm{NH}_3$
after sampling noise, truncation gap, and generalized-eigensolver instability
are taken into account. For the larger systems $\mathrm{HCN}$, $\mathrm{CO}_2$,
and $\mathrm{COCl}_2$, however, the present results indicate that further
progress is still needed. In those cases, the remaining limitations come partly
from the effect of sampling noise on the generalized eigensolver and partly
from the finite numerical resolution of weakly independent directions even in
the infinite-shot limit.

The main directions for further work are summarized below.

\begin{enumerate}[leftmargin=*,label=\arabic*.,itemsep=4pt,topsep=4pt]
\item \textbf{Design of the hybrid basis.} The effectiveness of CANOE depends on both 
the quality and the quantity of the added basis states. On the quantum side, one must 
determine which ansatz is most effective for representing the ground state. We used 
Krylov states here because they provide a simple and well-defined construction, but once 
the initial state is fixed they are largely deterministic, being generated by Hamiltonian 
time evolution alone. This convenience does not guarantee that they are the most effective 
choice for ground-state representation, and more flexible ans\"atze may perform 
better~\cite{Lee2019UpCCGSD,Fedorov2022}. On the classical side, both the construction and 
the optimal size of the determinant sector are also nontrivial, and alternative classical 
ans\"atze based on sampled determinants may also be useful to explore~\cite{Zhang2025}. 
As shown in \cref{appendix:nc_nq_sampled_eigensolve}, different combinations of $N_c$ and $N_q$ 
can lead to very different ground-state errors. In the infinite-shot limit, enlarging the basis 
should not worsen the variational solution. In the sampled setting, however, adding more 
directions can instead degrade the result because noise can outweigh the gain in expressivity 
and the strict variational character of the generalized eigenproblem may be lost.

\item \textbf{Plateau behavior and robust generalized eigensolvers.} The plateau behavior 
observed in the sampled-energy benchmarks is not yet fully understood. In our numerical tests, 
it appears to be related to the weight of the ground state on the nearly singular eigenvectors of 
the overlap matrix. This provides a useful consistency check, but it does not yet explain the 
plateau quantitatively, in particular its onset and magnitude across different systems. 
More generally, the eigensolver remains a key bottleneck for CANOE. In practice, it may fail 
to resolve subtle but physically meaningful differences between weakly independent quantum directions, 
thereby suppressing their contribution to the final solution. At the same time, sampled overlap 
matrices can become indefinite, so the noisy problem may leave the strictly variational regime. 
A more detailed analysis of how overlap-matrix noise propagates to the final energy error is therefore 
still needed, together with improved strategies that either enhance the numerical resolution of 
the generalized eigensolver in nearly singular hybrid subspaces or remain reliable when the 
sampled matrices are noisy and indefinite. Generalized eigenvalue solvers specialized for noisy matrices may solve some of these issues~\cite{hicks2023trimmed}.
\end{enumerate}

On the measurement side, it is also worth exploring whether
amplitude-estimation-based strategies, including recent Bayesian iterative
variants, can be adapted to the overlap-estimation problem to further reduce
sampling cost, although integrating such methods with the present histogram
protocol remains future work~\cite{Li2025BIQAE}.

If progress can be made on the two issues above, then a more meaningful
comparison with other early fault-tolerant quantum algorithms becomes
possible. Under the assumption that the plateau behavior can be mitigated by
improved treatment of the generalized eigensolver, the extrapolation in
\cref{appendix:shot_prediction} suggests that CANOE could still reach chemical
accuracy for larger systems at finite sampling cost.
Our current extrapolation gives a
per-histogram shot count on the order of $10^8$ at the 100-qubit scale, which
corresponds to a total classical--quantum-block sampling cost closer to
$10^{10}$ shots as the total shots for
the energy. This places CANOE in a
regime where comparison with VQE is most naturally made through total sampling
cost. Early resource estimates
for strongly correlated transition-metal chemistry gave VQE sampling costs as
large as $\sim 10^{13}$ shots per energy evaluation~\cite{Wecker2015}, while
later factorized measurement strategies reduce common bounds substantially, in
some cases by about three orders of magnitude~\cite{Huggins2021Meas}; however, the
total VQE cost still depends strongly on the ansatz, measurement scheme, and
number of optimization steps. By contrast, comparison with QPE and related
fault-tolerant quantum-simulation approaches is more
naturally made through coherent circuit cost. Recent local-control
formulations also suggest that phase-estimation-style algorithms can be adapted
toward shallower coherent implementations~\cite{Lin2022LectureNotes,Schiffer2025}.
In
\cref{appendix:error_rate_requirements}, a naive first-order Trotter estimate
with Jordan--Wigner decomposition gives a CANOE state-preparation depth of
order $10^{11}$ for a 100-qubit-scale system. Canonical Trotterized FeMoco QPE
estimates were on the order of $10^{14}$--$10^{15}$ logical non-Clifford
gates~\cite{Reiher2017}, whereas more optimized qubitization and
tensor-factorization analyses substantially reduce the fault-tolerant
cost, to around $10^9$ gates~\cite{low2025fast}.
These resource measures are not directly equivalent to our depth estimate, but
they suggest that CANOE may occupy an intermediate regime between
measurement-heavy variational methods and fully coherent phase-estimation
approaches.

From another perspective, one can treat CANOE as a way to push for a
wavefunction ansatz that is (under sufficient sampling conditions) strictly 
better than what can be achieved on a classical or a quantum computer alone.
Recent demonstrations of Fermi-Hubbard dynamics on a 72-qubit quantum
processor~\cite{alam2025programmabledigitalquantumsimulation} demonstrate
that time-evolution, even with Trotterization, is becoming possible. The 
states prepared in Ref.~\cite{alam2025programmabledigitalquantumsimulation} 
could be used to create the necessary quantum states. On the other side,
recent advances in selected configuration interaction methods have demonstrated
compact, expressive determinant expansions that can provide 
quality approximations of the ground state~\cite{Zhang2025}. CANOE could use these
state-of-the-art determinant expansions as the classical states. Putting them together,
with a sufficiently robust generalized eigensolver and sufficient shots, would result in an
estimate of the ground state that is better than either the classical or quantum computer
by themselves.

\section{Methods}
\label{sec:methods}
%
Here, we summarize the additional implementation details used in our numerical benchmarks, including the 
construction of the hybrid basis, the overlap-estimation workflows, and the generalized eigensolver.

\subsection{Hybrid-Basis Construction}
\label{subsec:method_basis_construction}

The construction of the classical and quantum basis states is described in \cref{subsec:subhyb};
here we note only the additional setup details used in practice. The Hamiltonian is constructed 
in the selected heat-bath configuration interaction (SHCI) variational space. The underlying determinant 
set is generated using a SHCI procedure~\cite{li2018fast} with variational selection threshold 
$\epsilon_{\mathrm{var}} = 5\times10^{-5}$. This yields a compact yet near-converged approximation to 
the full configuration interaction (FCI) space, enabling simulations of moderately large correlated systems. 
From this SHCI-selected determinant set, we retain the leading $N_c$ determinants ranked by variational 
coefficient. Within the same SHCI space, the quantum basis states are constructed to capture low-energy 
correlations. For the chromium numerical experiment in \cref{subsec:subhyb}, the SHCI variational 
space contains $1.4\times10^{6}$ determinants, from which we construct classical basis subsets of 
varying size $N_c$, together with up to 64 Krylov states as quantum basis states.

\subsection{Overlap estimation methods}

\subsubsection{Practical implementation of histogram-based estimation}
\label{subsec:histogram_practical_impl}

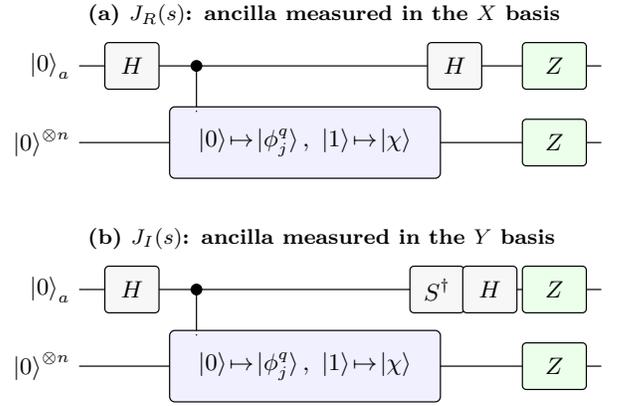
\begin{figure}[t]
\centering
\begin{tikzpicture}[
    x=0.78cm,
    y=0.85cm,
    line width=0.45pt,
    gate/.style={draw, rounded corners=1.5pt, minimum width=0.72cm, minimum height=0.62cm, fill=black!3},
    prep/.style={draw, rounded corners=2pt, minimum width=3.6cm, minimum height=0.95cm, fill=blue!6, align=center},
    meas/.style={draw, rounded corners=1.5pt, minimum width=0.85cm, minimum height=0.62cm, fill=green!8},
    lab/.style={font=\footnotesize}
]
    \node[anchor=west, font=\footnotesize\bfseries] at (0,3.2) {(a) $J_R(s)$: ancilla measured in the $X$ basis};
    \draw (0,2.4) node[left] {$\ket{0}_a$} -- (8.9,2.4);
    \draw (0,1.2) node[left] {$\ket{0}^{\otimes n}$} -- (8.9,1.2);

    \node[gate] at (0.9,2.4) {$H$};

    \filldraw (2.0,2.4) circle (2pt);
    \draw (2.0,2.4) -- (2.0,1.7);
    \draw (2.3,0.72) rectangle (5.4,1.68);
    \node[prep] at (3.85,1.2) {$\ket{0}\!\mapsto\!\ket{\phi_j^q},\ \ket{1}\!\mapsto\!\ket{\chi}$};
    \draw (2.0,1.7) -- (2.0,1.68);

    \node[gate] at (6.4,2.4) {$H$};
    \node[meas] at (8.1,2.4) {$Z$};
    \node[meas] at (8.1,1.2) {$Z$};

    \node[anchor=west, font=\footnotesize\bfseries] at (0,-0.3) {(b) $J_I(s)$: ancilla measured in the $Y$ basis};
    \draw (0,-1.1) node[left] {$\ket{0}_a$} -- (8.9,-1.1);
    \draw (0,-2.3) node[left] {$\ket{0}^{\otimes n}$} -- (8.9,-2.3);

    \node[gate] at (0.9,-1.1) {$H$};

    \filldraw (2.0,-1.1) circle (2pt);
    \draw (2.0,-1.1) -- (2.0,-1.8);
    \draw (2.3,-2.78) rectangle (5.4,-1.82);
    \node[prep] at (3.85,-2.3) {$\ket{0}\!\mapsto\!\ket{\phi_j^q},\ \ket{1}\!\mapsto\!\ket{\chi}$};
    \draw (2.0,-1.8) -- (2.0,-1.82);

    \node[gate] at (6.1,-1.1) {$S^\dagger$};
    \node[gate] at (7.0,-1.1) {$H$};
    \node[meas] at (8.1,-1.1) {$Z$};
    \node[meas] at (8.1,-2.3) {$Z$};
\end{tikzpicture}
\caption{Ancilla-assisted implementation of histogram-based overlap estimation. The ancilla-assisted preparation yields $\ket{\Psi}=\frac{1}{\sqrt{2}}(\ket{0}\ket{\phi_j^q}+\ket{1}\ket{\chi})$. Measuring the ancilla in the $X$ or $Y$ basis and the system register in the computational basis gives the joint histograms $J_R$ and $J_I$, respectively.}
\label{fig:histogram_circuit}
\end{figure}

Because the interference states $\ket{\psi_R}$ and $\ket{\psi_I}$ are not normalized, 
they cannot be prepared directly with a unitary circuit. Instead, we introduce an ancilla qubit and prepare
\begin{equation}
\ket{\Psi}:=\frac{1}{\sqrt{2}}\Big(\ket{0}\ket{\phi_j^q}+\ket{1}\ket{\chi}\Big).
\end{equation}
If the ancilla is measured in the $X$ basis, implemented by a final Hadamard gate followed by 
computational-basis readout, and the remaining qubits are measured in the $Z$ basis, then the 
joint histogram for determinant $s$ is
\begin{equation}\label{eq:histograms}
J_R(s):=\Pr(+_X,s)=\frac{1}{4}\,\big|\alpha_s+\beta_s\big|^2,
\end{equation}
where $\alpha_s=\braket{s|\phi_j^q}$ and $\beta_s=\braket{s|\chi}$. Replacing the final 
Hadamard by $HS^\dagger$, i.e., measuring the ancilla in the $Y$ basis, gives
\begin{equation}
J_I(s):=\Pr(+_Y,s)=\frac{1}{4}\,\big|\alpha_s+i\beta_s\big|^2.
\end{equation}
These histograms satisfy
\begin{equation}
p_R(s)=2J_R(s),\qquad p_I(s)=2J_I(s),
\end{equation}
so the estimator introduced in \cref{sec:results} can be evaluated directly from the measured joint 
histograms without any additional normalization of $\ket{\psi_R}$ and $\ket{\psi_I}$. 
A schematic circuit is shown in \cref{fig:histogram_circuit}.

In the numerical simulations reported here, we benchmarked histogram-based estimation across the 
molecular set $\mathrm{H}_4$, $\mathrm{H}_2\mathrm{O}$, $\mathrm{NH}_3$, $\mathrm{HCN}$, 
$\mathrm{CO}_2$, and $\mathrm{COCl}_2$. For each system, we first constructed an SHCI variational 
space with selection threshold $\epsilon_{\mathrm{var}}=10^{-5}$ and used the resulting variational 
energy as the reference ground-state energy within our numerical setup. This reference is not 
the exact FCI energy, but it is sufficiently accurate for our benchmarking purposes. The same 
SHCI space defines the determinant pool from which the CANOE basis is assembled. In the intended 
CANOE regime, we retain a large classical sector and a comparatively small number of quantum states, 
both defined within this SHCI space. To form the sampled overlap matrices, we use the identity for 
the classical--classical block and analytically computed values for the quantum--quantum block; 
in an actual quantum implementation, the latter would be obtained with the Hadamard test. 
Sampling noise is introduced only in the classical--quantum block, while the quantum--quantum 
block is kept at its analytically computed value.

\subsubsection{Practical implementation of classical shadow}

For local Pauli classical shadows, each quantum basis state is measured repeatedly in randomly 
chosen single-qubit Pauli bases. Each shot records both the basis choices and the corresponding 
bit string, from which a classical-shadow snapshot is reconstructed. To recover the complex 
overlap with a classical determinant $\ket{\phi_i^c}$, we define the off-diagonal observables
\begin{align*}
O_i^{(X)} &:= \ket{0^n}\bra{\phi_i^c} + \ket{\phi_i^c}\bra{0^n}, \\
O_i^{(Y)} &:= i\bigl(\ket{0^n}\bra{\phi_i^c} - \ket{\phi_i^c}\bra{0^n}\bigr),
\end{align*}
and prepare the reference-interference state
\begin{equation*}
\ket{\psi_R}=\frac{\ket{\phi_j^q}+\ket{0^n}}{\sqrt{2}}.
\end{equation*}
Their expectation values give the real and imaginary parts of the overlap,
\begin{align*}
\langle\psi_R|O_i^{(X)}|\psi_R\rangle
&= \operatorname{Re}\langle \phi_i^c|\phi_j^q\rangle, \\
\langle\psi_R|O_i^{(Y)}|\psi_R\rangle
&= \operatorname{Im}\langle \phi_i^c|\phi_j^q\rangle.
\end{align*}
In practice, this means that for each quantum basis state one must estimate a pair of observables 
for every classical determinant, i.e., $2N_c$ target observables in total. In our numerical 
implementation, we used the code from the ShadowGrouping reference~\cite{Gresch2025}, adjusted 
its parameters to implement randomized classical shadows, and interfaced it with our own 
routines to project the sampled states onto determinants in the SHCI space. More generally, 
software support for randomized-measurement workflows is also available through packages such as 
RandomMeas.jl~\cite{Elben2025RandomMeas}.

\subsection{Practical implementation of the generalized eigensolver}

For the numerical simulations in this work, the LOBPCG iterations were carried out with the 
implementation provided by \texttt{scipy.sparse.linalg.lobpcg}. This library routine was 
used within our own Schur-complement-based solver workflow to treat the generalized eigenvalue 
problem in the CANOE basis. For the sampled matrices, we use LOBPCG alone in a uniform manner 
across all systems. At moderate shot counts, the sampling noise effectively regularizes the 
ill-conditioning of $\mathcal{S}$, so additional deflation or pseudo-inverse stabilization is 
not yet necessary. As the sampling error is reduced and the calculation enters the high-accuracy 
regime, however, the underlying near-linear dependencies re-emerge, and the same deflation and 
pseudo-inverse treatment used for the infinite-shot limit becomes necessary. 
Its effect on the benchmarks is discussed in \cref{sec:results}.

\begin{acknowledgements}
We thank Preetham Tikkireddi for useful discussions. 
The authors acknowledge the use of OpenAI's ChatGPT and Codex 
for editorial assistance in organizing and refining the presentation of this manuscript.
This material is based upon work supported by NSF Award No.
2016136 for the QLCI center Hybrid Quantum Architectures and Networks 
and the U.S. Department of Energy Office of Science National Quantum 
Information Science Research Centers as part of the Q-NEXT center.
\end{acknowledgements}

%
\appendix
\onecolumngrid
\makeatletter
\def\section{%
  \@startsection
    {section}%
    {1}%
    {\z@}%
    {0.8cm \@plus1ex \@minus .2ex}%
    {0.5cm}%
    {\normalfont\normalsize\bfseries\centering}%
}%
\makeatother
\renewcommand{\figurename}{Supplementary Figure}
\setcounter{figure}{0}
\renewcommand{\thefigure}{S\arabic{figure}}
\renewcommand{\theHfigure}{supp.\arabic{figure}}
%
\clearpage
\section{Sample-complexity derivations for overlap-estimation methods}
\label{appendix:sample_complexity}

\subsection{Hadamard test}

To estimate an overlap or Hamiltonian matrix element between two quantum
basis states $\ket{\phi_i^q}$ and $\ket{\phi_j^q}$, the Hadamard-test circuit
creates an ancilla-mediated interference state. Following
Ref.~\cite{Ren2025}, the final state can be written as
\begin{align}
\ket{\phi_{ij}^{\mathrm{end}}}
= \frac{1}{\sqrt{2}} \left( \ket{\phi_i^q} \ket{+} + e^{i\theta} \ket{\phi_j^q} \ket{-} \right).
\end{align}
Measuring $I\otimes Z$ on this state yields the overlap matrix element. For $\theta=0$,
\begin{align}
\bra{\phi_{ij}^{\mathrm{end}}} I \otimes Z \ket{\phi_{ij}^{\mathrm{end}}}
= \mathrm{Re}\!\left( \braket{\phi_i^q | \phi_j^q} \right)
= \mathrm{Re}(S_{ij}),
\end{align}
while for $\theta=\pi/2$,
\begin{align}
\bra{\phi_{ij}^{\mathrm{end}}} I \otimes Z \ket{\phi_{ij}^{\mathrm{end}}}
= \mathrm{Im}\!\left( \braket{\phi_i^q | \phi_j^q} \right)
= \mathrm{Im}(S_{ij}).
\end{align}

Let
\begin{equation}
H = \sum_{k=1}^{n_H} h_k P_k.
\end{equation}
Then the same construction gives the Hamiltonian matrix element. For $\theta=0$,
\begin{align}
\bra{\phi_{ij}^{\mathrm{end}}} H \otimes Z \ket{\phi_{ij}^{\mathrm{end}}}
= \mathrm{Re}\!\left( \bra{\phi_i^q} H \ket{\phi_j^q} \right)
= \mathrm{Re}(H_{ij}),
\end{align}
while for $\theta=\pi/2$,
\begin{align}
\bra{\phi_{ij}^{\mathrm{end}}} H \otimes Z \ket{\phi_{ij}^{\mathrm{end}}}
= \mathrm{Im}\!\left( \bra{\phi_i^q} H \ket{\phi_j^q} \right)
= \mathrm{Im}(H_{ij}).
\end{align}

Let $U_{\mathrm{end}}$ denote the circuit that prepares
$\ket{\phi_{ij}^{\mathrm{end}}}$. Using the Pauli decomposition, the
real-part estimator can be written as
\begin{align}
\bra{\phi_{ij}^{\mathrm{end}}} H \otimes Z \ket{\phi_{ij}^{\mathrm{end}}}
=\sum_{k=1}^{n_H}h_k\braket{0|U_{\mathrm{end}}^{\dagger}(P_k \otimes Z) U_{\mathrm{end}}|0}.
\end{align}
Assume that the $n_H$ Pauli terms are measured independently and that
each term is assigned the same number of shots $n_{\mathrm{meas}}$. Let
$X\in\{-1,+1\}$ denote the outcome of a single Pauli measurement, with
probabilities $p$ and $q=1-p$. Then
\begin{equation}
\mu=\mathbb{E}[X]=p-(1-p)=2p-1,
\end{equation}
and the variance of the estimated probability is
\begin{equation}
\operatorname{Var}(\hat{p})=\frac{pq}{n_{\mathrm{meas}}}.
\end{equation}
Therefore, the variance of the sample mean of $X$ satisfies
\begin{equation}
\operatorname{Var}(\hat{X})=4\operatorname{Var}(\hat{p})=\frac{4pq}{n_{\mathrm{meas}}}\leq\frac{1}{n_{\mathrm{meas}}}.
\end{equation}
The variance of the estimator for
$\bra{\phi_{ij}^{\mathrm{end}}} H \otimes Z \ket{\phi_{ij}^{\mathrm{end}}}$ is then bounded by
\begin{equation}
\operatorname{Var}\leq\sum_{k=1}^{n_H} h_k^2 \operatorname{Var}(\hat{\mu}_k)
\leq\sum_{k=1}^{n_H}\frac{h_k^2}{n_{\mathrm{meas}}}
=\frac{\|\mathbf{H}\|_2^2}{n_{\mathrm{meas}}},
\end{equation}
where $n_{\mathrm{meas}}$ is the number of shots per Pauli observable.

If the target standard error for $\mathrm{Re}(H_{ij})$ is $\epsilon$,
it is sufficient to choose
\begin{equation}
n_{\mathrm{meas}}\ge \frac{\|\mathbf{H}\|_2^2}{\epsilon^2}.
\end{equation}
Because the $n_H$ observables are measured separately, the total number
of shots sufficient to estimate one real-valued matrix element is
\begin{equation}
n_{\mathrm{shot}}^{\mathrm{Re}(H_{ij})}
\ge n_H\,n_{\mathrm{meas}}
\ge \frac{n_H\|\mathbf{H}\|_2^2}{\epsilon^2}.
\end{equation}
Consequently, the number of shots required for the full
$N_c\times N_q$ classical-quantum block of the Hamiltonian matrix is
\begin{equation}
\boxed{
n_{\mathrm{block}_{cq}}
\ge 2N_c N_q\, n_{\mathrm{shot}}^{\mathrm{Re}(H_{ij})}
\ge \frac{2N_c N_q n_H\|\mathbf{H}\|_2^2}{\epsilon^2}
},
\end{equation}
where the factor of $2$ accounts for measuring both the real and imaginary parts.

\subsection{Classical shadow}
\label{appendix:classical_shadow}

To recover the complex overlap $\langle \phi_i^c | \phi_j^q \rangle$
with classical shadows, it is convenient to introduce observables whose
expectation values directly return its real and imaginary parts with
respect to the reference string $\ket{0^n}$. For classical-shadow
estimation, define the off-diagonal observables
\begin{align}
O_i^{(X)} &:= \ket{0^n}\bra{\phi_i^c} + \ket{\phi_i^c}\bra{0^n}, \\
O_i^{(Y)} &:= i\bigl(\ket{0^n}\bra{\phi_i^c} - \ket{\phi_i^c}\bra{0^n}\bigr).
\end{align}
Using the same reference-interference construction, for a target
quantum basis state $\ket{\phi_j^q}$,
consider the state
\begin{align}
\ket{\psi_R}=\frac{\ket{\phi_j^q}+\ket{0^n}}{\sqrt{2}},
\end{align}
the expectation values of these observables satisfy
\begin{align}
\langle\psi_R|O_i^{(X)}|\psi_R\rangle
&= \tfrac12\bigl(\langle\phi_j^q|\phi_i^c\rangle+\langle \phi_i^c|\phi_j^q\rangle\bigr)
= \operatorname{Re}\langle \phi_i^c|\phi_j^q\rangle, \\
\langle\psi_R|O_i^{(Y)}|\psi_R\rangle
&= \tfrac12\bigl(-i\langle\phi_j^q|\phi_i^c\rangle + i\langle \phi_i^c|\phi_j^q\rangle\bigr)
= \operatorname{Im}\langle \phi_i^c|\phi_j^q\rangle.
\end{align}
Therefore, by preparing the corresponding state $\ket{\psi_R}$ for each
quantum basis state $\ket{\phi_j^q}$ and estimating $O_i^{(X)}$ and
$O_i^{(Y)}$ with classical shadows, one can reconstruct the complex
overlap element $S_{ij}$.

For each quantum state, this requires estimating the two observables
$O_i^{(X)}$ and $O_i^{(Y)}$ for every classical basis state $\ket{\phi_i^c}$, so the total
number of target observables is $M=2N_c$. Using the standard
classical-shadow bound for simultaneously estimating $M$ observables,
the required number of shots is
\begin{equation}
n_{\mathrm{shot}}
= KN
= 2\log\!\left(\frac{2M}{\delta}\right)
\times
\frac{34}{\epsilon^2}
\max_{1\le i\le M}\left\|O_i\right\|_{\mathrm{shadow}}^2.
\end{equation}
In the present case, the relevant observables are off-diagonal many-body
operators. For local Pauli classical shadows, a worst-case bound is
\[
\|O_i\|_{\mathrm{shadow}}^2=\mathrm{O}(3^{n_{\mathrm{qubit}}}).
\]
Using $M=2N_c$, this gives the sufficient bound
for the overlap estimation of one quantum basis state $\ket{\phi_j^q}$,
where $\eta$ is the target overlap error. Since $H_{ij}$ is reconstructed from the sampled
overlaps through \cref{eq:H_ij}, the induced error satisfies the
sufficient bound
\begin{equation}
|\,\widehat{H}_{ij}-H_{ij}\,|
\le
\lVert\mathbf{H}\rVert_{1}\eta.
\end{equation}
Therefore, to achieve a target error $\epsilon$ in
$\mathrm{Re}(H_{ij})$, it is sufficient to choose
\begin{equation}
n_{\mathrm{shot}}^{\mathrm{Re}(H_{ij})}
\ge
68\,3^{n_{\mathrm{qubit}}}
\frac{\lVert\mathbf{H}\rVert_{1}^2
\log\!\left(4N_c/\delta\right)}{\epsilon^2}.
\end{equation}
Across all $N_q$ quantum basis states, the full classical--quantum
block then requires
\begin{equation}
\boxed{
n_{\mathrm{block}_{cq}}
\ge
N_q\,n_{\mathrm{shot}}^{\mathrm{Re}(H_{ij})}
\ge
68\,N_q\,3^{n_{\mathrm{qubit}}}
\frac{\lVert\mathbf{H}\rVert_{1}^2
\log\!\left(4N_c/\delta\right)}{\epsilon^2},
}
\end{equation}
Note that this estimate assumes local Pauli classical shadows; other shadow protocols may exhibit different sample-complexity scalings.
Table~II lists only the corresponding big-$O$ scaling.

\subsection{Histogram-based estimation}

Fix one quantum basis state $\ket{\phi_j^q}$. Partition the selected classical
strings into $B$ disjoint batches of size $m$, so
$B=\lceil N_c/m\rceil$. For batch
$k$, define the reference state
\begin{equation}
\ket{\chi_k}=\frac{1}{\sqrt{m}}\sum_{s\in S_k}\ket{s},
\end{equation}
so that
\begin{equation}
\beta_s\equiv\braket{s|\chi_k}=
\begin{cases}
1/\sqrt{m}, & s\in S_k,\\
0, & s\notin S_k.
\end{cases}
\end{equation}
For each quantum basis state, one measures one reference histogram and
$2B$ interference histograms, giving $1+2B$ histograms in total. Over
all $N_q$ quantum states, this becomes $N_q(1+2B)$ histograms. Let
$n_{\mathrm{meas}}$ denote the number of shots per histogram.
Here $n_{\mathrm{meas}}$ denotes the shots used to estimate a single
histogram, whereas $n_{\mathrm{shot}}$ below denotes the total shot
count summed over all histograms and all quantum basis states.
\newline
\paragraph{Estimator.}
For $s\in S_k$, where $\beta_s=1/\sqrt{m}$, the estimator is
\begin{align}
\hat\alpha_s
=&
\frac{\hat p_R^{(k)}(s)-\tfrac12\big(\hat p_q(s)+|\beta_s|^2\big)+i\big[\hat p_I^{(k)}(s)-\tfrac12\big(\hat p_q(s)+|\beta_s|^2\big)\big]}
{\overline{\beta_s}}\\
=&
\sqrt{m}\,\Big[\hat p_R^{(k)}(s)-\tfrac12\big(\hat p_q(s)+\tfrac1m\big)\Big]
+
i\sqrt{m}\,\Big[\hat p_I^{(k)}(s)-\tfrac12\big(\hat p_q(s)+\tfrac1m\big)\Big].
\end{align}

\paragraph{Hoeffding bounds for histogram errors.}
For a full histogram on $n_{\mathrm{qubit}}$ qubits, there are
$2^{n_{\mathrm{qubit}}}$ computational-basis outcomes. Applying
Hoeffding's inequality to each bin and a union bound over all bins gives
\begin{equation}
\Pr\!\left(\max_{s}|\hat p(s)-p(s)|\ge \varepsilon\right)
\le 2\cdot 2^{n_{\mathrm{qubit}}}e^{-2n_{\mathrm{meas}}\varepsilon^2}.
\end{equation}
Applying a union bound over the $N_q$ reference histograms and the
$N_qB$ real and imaginary interference histograms gives
\begin{equation}
\begin{aligned}
&\Pr\;\!\Bigl(
\max\{|\Delta p_q|,|\Delta p_R|,|\Delta p_I|\}
\ge \varepsilon
\Bigr)
\le
2\cdot 2^{n_{\mathrm{qubit}}}N_q(1+2B)\,
e^{-2n_{\mathrm{meas}}\varepsilon^2}.
\end{aligned}
\end{equation}
Thus, with confidence $1-\delta$,
\begin{equation}
\varepsilon
\le
\sqrt{\frac{1}{2n_{\mathrm{meas}}}
\ln\!\frac{2\cdot 2^{n_{\mathrm{qubit}}}N_q(1+2B)}{\delta}}.
\label{eq:histogram_eps}
\end{equation}

\paragraph{Propagation to amplitude error.}
If $|\Delta p_q|\le \varepsilon$, $|\Delta p_R|\le \varepsilon$, and
$|\Delta p_I|\le \varepsilon$, then for $s\in S_k$ one has
\begin{align}
|\Re(\hat\alpha_s-\alpha_s)|
&=
\frac{\big|(\Delta p_R)-\tfrac12(\Delta p_q)\big|}{|\beta_s|}
\le \tfrac32\sqrt{m}\,\varepsilon,\\
|\Im(\hat\alpha_s-\alpha_s)|
&=
\frac{\big|(\Delta p_I)-\tfrac12(\Delta p_q)\big|}{|\beta_s|}
\le \tfrac32\sqrt{m}\,\varepsilon.
\end{align}
Therefore,
\begin{equation}
|\hat\alpha_s-\alpha_s|
\le
\sqrt{2}\cdot\tfrac32\sqrt{m}\,\varepsilon
\le
\frac{3}{2}\sqrt{\frac{m}{n_{\mathrm{meas}}}
\ln\!\frac{2\cdot 2^{n_{\mathrm{qubit}}}N_q(1+2B)}{\delta}}.
\label{eq:histogram_alpha_error}
\end{equation}

\paragraph{Per-amplitude target accuracy.}
Demanding $|\hat\alpha_s-\alpha_s|\le \eta$ uniformly over all strings
and all quantum states with confidence $1-\delta$ is guaranteed if
\begin{equation}
n_{\mathrm{meas}}
\ge
\frac{9}{4}\,\frac{m}{\eta^2}
\ln\!\frac{2\cdot 2^{n_{\mathrm{qubit}}}N_q(1+2B)}{\delta}.
\label{eq:histogram_N_req}
\end{equation}

\paragraph{Total shot count.}
Each quantum basis state requires $(1+2B)n_{\mathrm{meas}}$ shots, so over all $N_q$ quantum
states
\begin{equation}
n_{\mathrm{shot}}
\ge
N_q(1+2B)
\frac{9}{4}\,\frac{m}{\eta^2}
\ln\!\left(
\frac{2\cdot 2^{n_{\mathrm{qubit}}}N_q(1+2B)}{\delta}
\right).
\end{equation}
Finally, since the error in
$\langle s|H|\phi_j^q\rangle
=\sum_{\gamma} h_{\gamma}(-i)^{|Y_{\gamma}|}(-1)^{s\cdot p_{\gamma}}
\alpha_{s\oplus b_{\gamma}}$
scales as $\lVert\mathbf{H}\rVert_{1}\eta$, setting
$\epsilon=\lVert\mathbf{H}\rVert_{1}\eta$ gives
\begin{equation}
\boxed{
n_{\mathrm{shot}}
\ge
N_q(1+2B)\,
\frac{9}{4}\,\frac{m\lVert\mathbf{H}\rVert_{1}^2}{\epsilon^2}
\ln\!\left(
\frac{2\cdot 2^{n_{\mathrm{qubit}}}N_q(1+2B)}{\delta}
\right).
}
\end{equation}
Table~II lists only the corresponding big-$O$ scaling.

These sample-complexity bounds quantify the measurement effort required
to control the reconstruction error of the sampled overlap and
Hamiltonian matrices. A convenient perturbative regime is
\begin{equation}
\|\delta H\|_2 + |E_0|\,\|\delta S\|_2 \ll \Delta_{\mathrm{gen}},
\qquad
\Delta_{\mathrm{gen}}:=\min_{k\ne 0}|E_k-E_0|,
\end{equation}
where $E_k$ are the exact generalized eigenvalues. In this regime,
first-order perturbation theory can be used to estimate the induced
ground-state energy error. Bounds on the sampled errors in $H$ and $S$
then translate directly into a bound on the energy shift through the
first-order relation
\[
\delta E^{(1)} = v_0^{\dagger}\!\bigl(\delta H - E_0\,\delta S\bigr)v_0.
\]
Coherent hardware noise or correlated readout errors
are not included in this statistical model; such effects can introduce a
bias term and modify the prefactor of the energy-error bound, but do not
change the leading $n_{\mathrm{shot}}^{-1/2}$ scaling of the sampling
error to first order.

\clearpage
\section{Reconstructing the quantum--quantum block from determinant overlaps}
\label{appendix:qq_from_overlaps}

Let
\begin{equation}
\alpha_s^{(j)}:=\braket{s|\phi_j^q},
\qquad
\alpha_s^{(k)}:=\braket{s|\phi_k^q},
\end{equation}
denote the determinant amplitudes of two quantum basis states
$\ket{\phi_j^q}$ and $\ket{\phi_k^q}$. If these amplitudes are reconstructed
on a common determinant set $\mathcal{D}$ and with a consistent phase
convention, then the quantum--quantum overlap element can be written as
\begin{equation}
\mathcal{S}_{jk}
=
\braket{\phi_j^q|\phi_k^q}
=
\sum_{s} \overline{\alpha_s^{(j)}}\,\alpha_s^{(k)}.
\end{equation}
Therefore, when the determinant amplitudes are known on the full support of
the two states, the overlap block $\mathcal{S}_{qq}$ can be assembled
classically. If only the amplitudes on a restricted set $\mathcal{D}$ are
available, the corresponding truncated approximation is
\begin{equation}
\widetilde{\mathcal{S}}_{jk}^{(\mathcal{D})}
=
\sum_{s\in\mathcal{D}}
\overline{\alpha_s^{(j)}}\,\alpha_s^{(k)}.
\end{equation}
In general, $\widetilde{\mathcal{S}}_{jk}^{(\mathcal{D})}$ is not exact, since
contributions from determinants outside $\mathcal{D}$ are omitted. The special
case $j=k$ becomes exact only when $\mathcal{D}$ is taken to be the full
computational-basis support of the same quantum state, so that the sum reduces
to the full histogram normalization
$\sum_s |\alpha_s^{(j)}|^2=1$. For a restricted selected determinant set, even
the diagonal case remains a truncated approximation.

For the Hamiltonian block, using the Pauli decomposition
\begin{equation}
H=\sum_{\gamma=1}^{n_H} h_{\gamma}P_{\gamma},
\end{equation}
and the computational-basis action
\begin{equation}
P_{\gamma}\ket{s}=\theta_{\gamma}(s)\ket{s\oplus b_{\gamma}},
\end{equation}
one obtains
\begin{align}
\mathcal{H}_{jk}
&=
\bra{\phi_j^q}H\ket{\phi_k^q}
=
\sum_{\gamma=1}^{n_H} h_{\gamma}
\bra{\phi_j^q}P_{\gamma}\ket{\phi_k^q}
=
\sum_{\gamma=1}^{n_H} h_{\gamma}
\sum_s
\overline{\alpha_s^{(j)}}\,\theta_{\gamma}(s)\,
\alpha_{s\oplus b_{\gamma}}^{(k)}.
\label{eq:qq_from_overlaps}
\end{align}
This shows that off-diagonal quantum--quantum Hamiltonian elements are also
recoverable in principle from determinant amplitudes, not only the diagonal
case $j=k$.

If amplitudes are available only on a restricted determinant set
$\mathcal{D}$, then for each Pauli term one must also require the shifted
string $s\oplus b_{\gamma}$ to remain inside $\mathcal{D}$. Defining
\begin{equation}
\mathcal{D}_{\gamma}
:=
\{\,s\in\mathcal{D}\;|\;s\oplus b_{\gamma}\in\mathcal{D}\,\},
\end{equation}
the corresponding truncated reconstruction is
\begin{equation}
\widetilde{\mathcal{H}}_{jk}^{(\mathcal{D})}
=
\sum_{\gamma=1}^{n_H} h_{\gamma}
\sum_{s\in\mathcal{D}_{\gamma}}
\overline{\alpha_s^{(j)}}\,\theta_{\gamma}(s)\,
\alpha_{s\oplus b_{\gamma}}^{(k)}.
\end{equation}
Hence the reconstruction is exact only when the chosen determinant set is
sufficiently complete for all Pauli-shifted amplitudes that contribute
appreciably to \cref{eq:qq_from_overlaps}. When that condition is not
satisfied, direct interference-based measurements such as the Hadamard test
remain a more straightforward way to evaluate the quantum--quantum block.

\clearpage
\section{Dependence of sampled CANOE simulations on \texorpdfstring{$N_c$}{Nc} and \texorpdfstring{$N_q$}{Nq}}
\label{appendix:nc_nq_sampled_eigensolve}

In the infinite-shot limit, enlarging the hybrid basis should not
worsen the variational ground-state energy; this behavior is consistent
with the discussion in Sec.~II. Once sampling noise is introduced,
however, the generalized eigenproblem need not remain strictly
variational. In that regime, adding additional classical or quantum
directions can amplify the effect of noise and increase the final energy
error rather than improve the solution. To examine this behavior
systematically, we varied the numbers of classical and quantum states,
$N_c$ and $N_q$, and computed the resulting ground-state errors for all
benchmark molecules.

For each molecule, we evaluated the ground-state energy error on a grid
of truncated classical and quantum subspace sizes $(N_c,N_q)$ at fixed
shot counts of $10^3$ and $10^7$, as shown in the two figures below. For
each case, we formed the sampled generalized eigenvalue problem
\[
H x = E S x,
\qquad
S=
\begin{pmatrix}
I & S_{cq}\\
S_{cq}^{\dagger} & S_{qq}
\end{pmatrix},
\qquad
H=
\begin{pmatrix}
H_{cc} & H_{cq}\\
H_{cq}^{\dagger} & H_{qq}
\end{pmatrix},
\]
where the required blocks were obtained by truncating the stored
sampled matrices to the first $N_c$ classical states and first $N_q$
quantum states. The assembled matrices were Hermitized before solution,
and the lowest generalized eigenvalue was computed with the LOBPCG
solver using $\mathrm{tol}=10^{-8}$ and $\mathrm{maxiter}=300$, with
multiple random complex initial vectors for robustness. The resulting
sampled ground-state energy $E_{\mathrm{sampled}}$ was compared against
the precomputed reference energy $E_{\mathrm{true}}$ (ground state energy
from SHCI), and each heatmap
entry reports the mean value of
$|E_{\mathrm{sampled}}-E_{\mathrm{true}}|$ over independent random-seed
realizations. The cell colors use a logarithmic scale, and the printed
values are the corresponding mean absolute errors in scientific
notation.

The resulting heatmaps show a clear difference between the low- and
high-shot regimes. At the lower shot count,
$n_{\mathrm{shot}}=10^3$, increasing $N_c$ or $N_q$ does not
systematically improve the ground-state energy error, and larger
subspaces often provide little benefit or even worsen the result. In
this regime, sampling noise is large enough that the sampled
generalized eigenproblem no longer cleanly reflects the variational
advantage of additional basis directions. At the higher shot count,
$n_{\mathrm{shot}}=10^7$, the benefit of enlarging the hybrid basis
becomes more visible, especially for the smaller systems, where lower
error regions extend more clearly toward larger $N_c$ and $N_q$. The
larger systems also improve relative to the $10^3$-shot case, but the
trend remains less pronounced, indicating that noise is reduced yet
still significant enough to mask the full advantage of additional
classical and quantum directions. This indicates that reducing the
ground-state energy error depends not only on the number of retained
states, but also on the quality of those states as an effective
variational basis in the presence of sampling noise.

\clearpage
\refstepcounter{figure}\label{fig:appendix_nc_nq_1e3}
\begin{center}
\includegraphics[height=0.41\textheight,keepaspectratio]{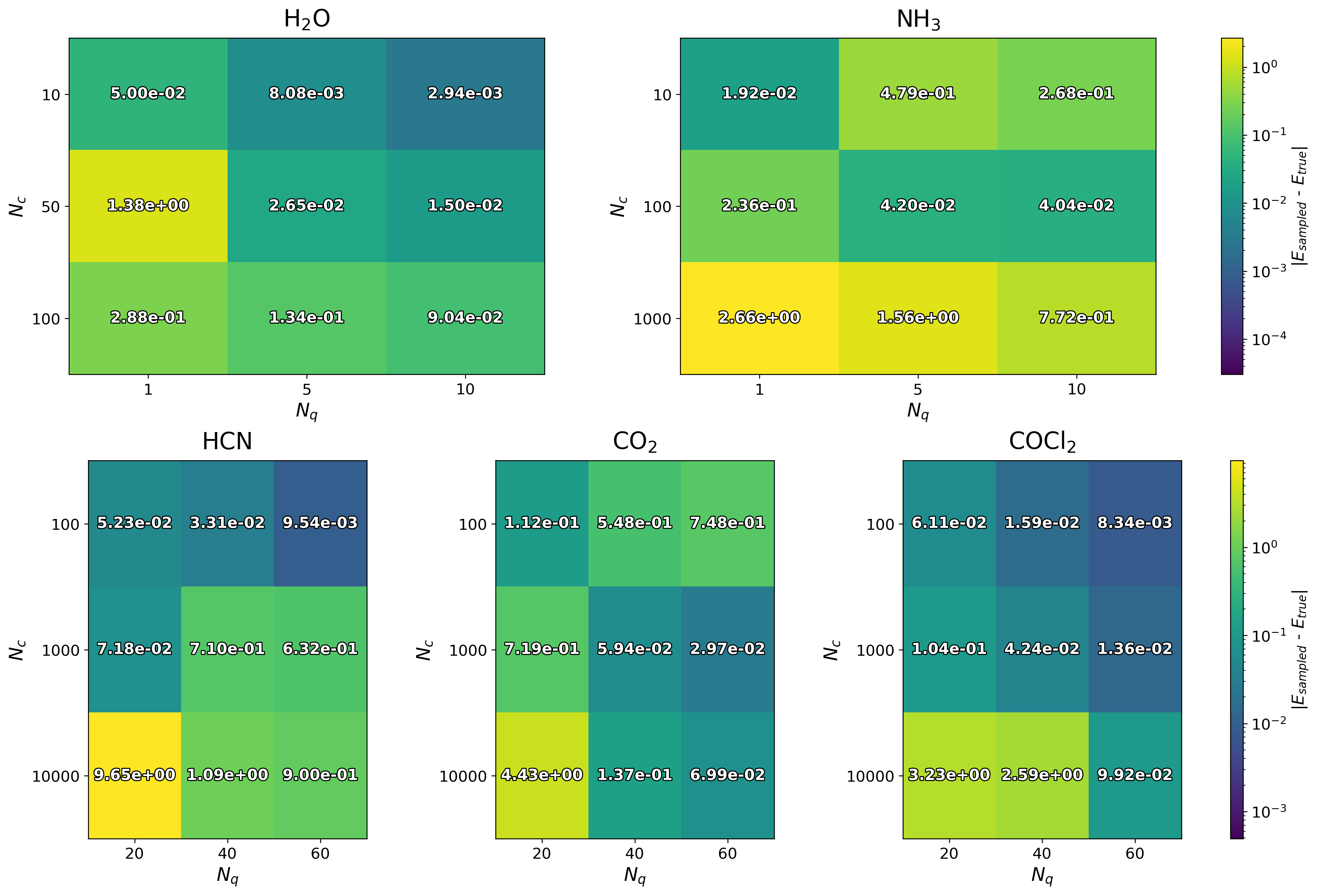}

{\small \figurename~\thefigure. Ground-state energy error over the
$(N_c,N_q)$ grid at fixed shot count $n_{\mathrm{shot}}=10^3$.}
\end{center}

\vspace{0.5em}
\refstepcounter{figure}\label{fig:appendix_nc_nq_1e7}
\begin{center}
\includegraphics[height=0.41\textheight,keepaspectratio]{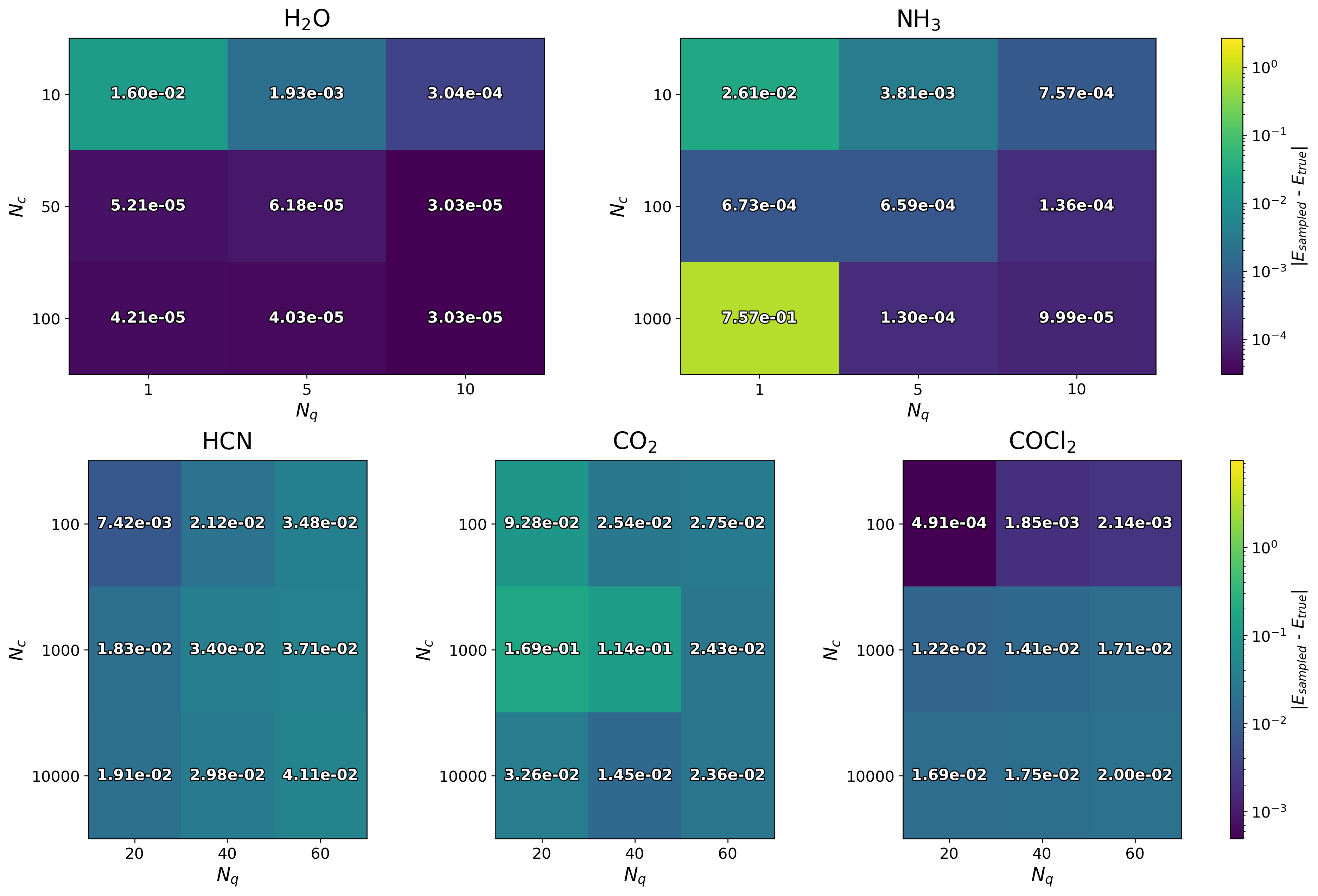}

{\small \figurename~\thefigure. Ground-state energy error over the
$(N_c,N_q)$ grid at fixed shot count $n_{\mathrm{shot}}=10^7$.}
\end{center}

\clearpage
\section{Threshold dependence of Schur-complement eigensolvers for sampled matrices}
\label{appendix:threshold_dependence}

\begin{figure}[H]
\centering
\includegraphics[height=0.78\textheight,keepaspectratio]{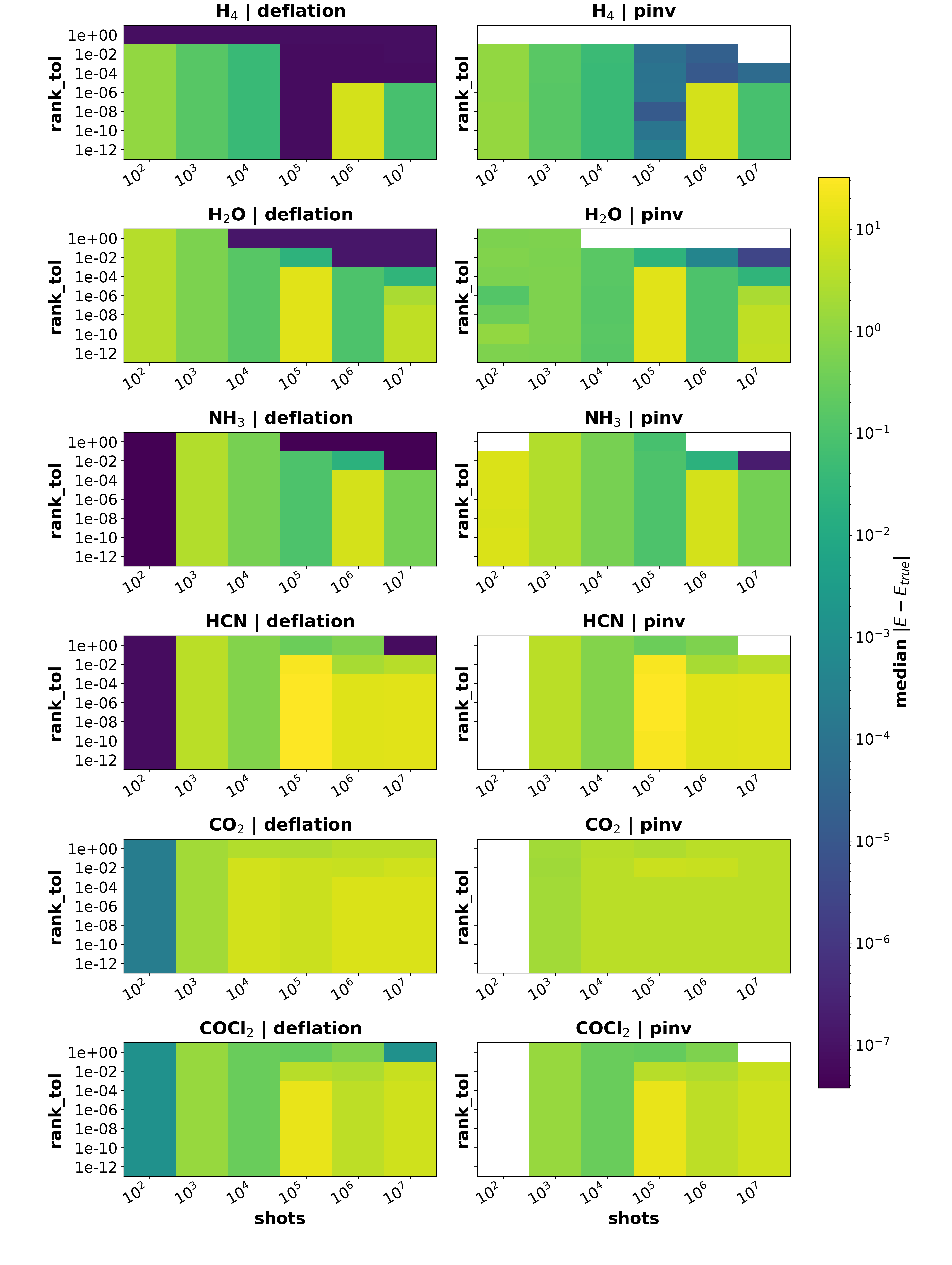}
\caption{Threshold dependence of sampled-matrix Schur-complement eigensolvers.
Rows correspond to benchmark molecules, columns to deflation and pseudo-inverse
regularization, and color to the median absolute ground-state energy error. White cells indicate parameter settings where the pinv Schur-complement solver returned no finite value because all Schur-complement directions were removed by the threshold.}
\label{fig:appendix_threshold}
\end{figure}

\clearpage

For each molecule, shot count, and random seed, we formed the sampled overlap
matrix and built the corresponding Schur-complement block used by the
deflation and pseudo-inverse solvers. We then swept the threshold parameter
\texttt{rank\_tol} over the values shown on the vertical axis and solved the
thresholded generalized eigenvalue problem with either deflation or
pseudo-inverse regularization. For each threshold, this gives a ground-state
energy estimate, which we compare with the SHCI reference energy. The color
plotted in \cref{fig:appendix_threshold} is the median absolute energy error over seeds, so each heatmap shows how
the solver accuracy changes jointly with shot count and threshold. The
horizontal axis is the number of shots, the vertical axis is the chosen
threshold, and the two columns compare deflation with pseudo-inverse
regularization.

\clearpage

\section{Shot-count extrapolation with qubit number}
\label{appendix:shot_prediction}

\begin{figure}[H]
\centering
\includegraphics[height=0.31\textheight,keepaspectratio]{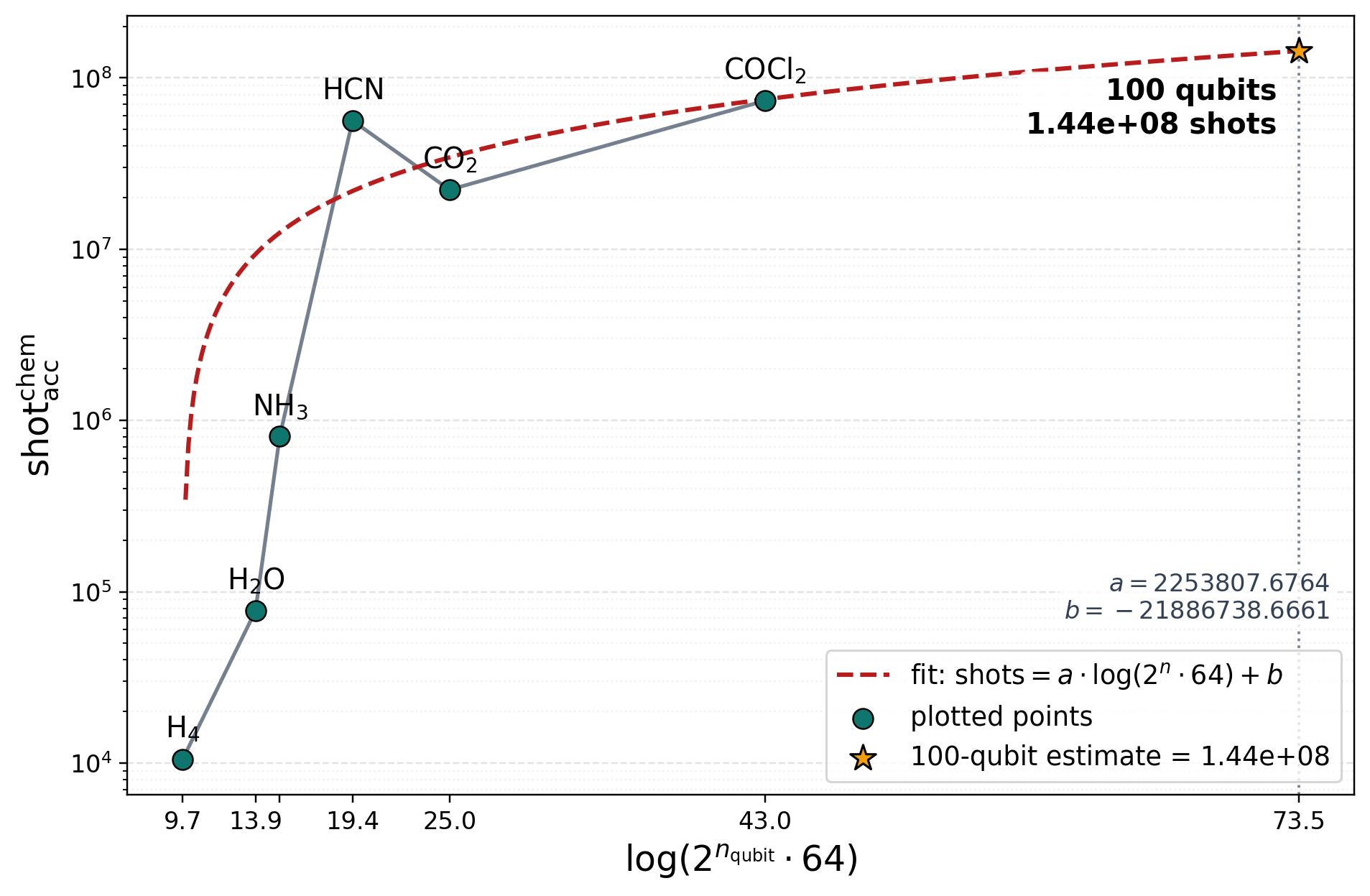}
\caption{Normalized chemical-accuracy shot cost versus
$\log(2^{n_{\mathrm{qubit}}}\cdot 64)$. Points are the adjusted per-histogram
shot costs, the solid line is the fitted model, and the star marks the
extrapolated 100-qubit estimate.}
\label{fig:shots-fit100-nq64-normalized}
\end{figure}

\Cref{fig:shots-fit100-nq64-normalized} was constructed from the
per-molecule energy-error curves shown in of
\cref{fig:overlap_benchmark}(d). For each molecule, we considered the curve of
mean absolute energy error, $|\Delta E|$, versus shot count and formed its
lower envelope, so that only the best achieved error up to each shot level was
retained. Chemical accuracy was defined as
$1.5936\times 10^{-3}\,\mathrm{Ha}$, and the corresponding shot cost,
$\mathrm{shot}_{\mathrm{acc}}^{\mathrm{chem}}$, was estimated as the shot count
at which the lower-envelope curve reaches this threshold. Here, the shot count
is the per-histogram shot count appearing on the horizontal axis of 
\cref{fig:overlap_benchmark}(d). The corresponding total cost for the full
classical--quantum block is obtained by multiplying this value by
$N_q(1+2B)$, since each quantum basis state requires one reference histogram
and $2B$ interference histograms. When the threshold was crossed within the
measured shot range, the crossing point
was obtained by log-log interpolation between neighboring data points. When the
threshold was not reached, the shot cost was estimated by log-log extrapolation
of the lower-envelope trend. This extrapolation assumes that any plateau in
the histogram-sampling curve, with shot count on the horizontal axis and energy
error on the vertical axis, is eventually improved at higher shot count so that
the lower-envelope error continues to decrease.

To compare molecules on the same theoretical footing, these shot costs were
then normalized to a common $N_q=64$ using
\[
\mathrm{shots}_{\mathrm{adj}}
=
\mathrm{shots}_{\mathrm{orig}}
\frac{\log(2^{n_{\mathrm{qubit}}}\cdot 64)}
{\log(2^{n_{\mathrm{qubit}}}\cdot N_q^{\mathrm{current}})}.
\]
After this normalization, we plotted the adjusted chemical-accuracy shot cost
against the theoretical variable $\log(2^{n_{\mathrm{qubit}}}\cdot 64)$ and fit
the data with the model
\[
\mathrm{shot}_{\mathrm{acc}}^{\mathrm{chem}}
=
a\,\log(2^{n_{\mathrm{qubit}}}\cdot 64)+b.
\]
This fit is applied at the per-quantum-state, per-histogram sample-complexity
level, before converting to the total classical--quantum-block cost through the
factor $N_q(1+2B)$. In this corrected model,
the 100-qubit extrapolation is approximately $1.44\times 10^8$ shots per
histogram. Multiplying by $N_q(1+2B)$ then gives a total classical--quantum
block cost of order $10^{10}$ shots. This estimate is based on the present
workflow and does not include possible further reductions from more advanced
factorization or compression strategies.

\clearpage

\section{Error-rate requirements for CANOE state preparation}
\label{appendix:error_rate_requirements}

CANOE can be implemented at either the physical-qubit or logical-qubit level,
so the error rate $p$ may denote either the physical error rate
$p_{\mathrm{phys}}$ or the logical error rate $p_{\mathrm{logic}}$. Let
$\delta$ denote the allowed failure probability for a single sampling run.
Under a worst-case assumption, the required error rate is bounded by
\[
\delta \ge 1-(1-p)^{D_{\mathrm{gate}}},
\]
where $D_{\mathrm{gate}}$ is the circuit gate depth. Assuming that the
coherence time is much longer than the gate time, decoherence can be neglected
and $p$ can be interpreted as a gate error rate.

The quantum states prepared in CANOE are Krylov states
\[
\ket{q_k} = \left(e^{-iH\Delta t}\right)^k \ket{q_0},
\]
which can be first-order Trotterized as
\[
\ket{q_k}
\approx
\left(\prod_{\ell=1}^{L} e^{-iH_\ell \delta t}\right)^{kr} \ket{q_0},
\]
where $H_\ell = h_\ell P_\ell$. In a naive second-quantized formulation,
$L\sim n_{\mathrm{qubit}}^4$, so
\[
D_{\mathrm{gate}} \approx Lrk.
\]
For a FeMoco-scale estimate~\cite{Montgomery2018FeMoco,Li2019FeMoco,Reiher2017,Lee2021THC},
taking $k=64$, $n_{\mathrm{qubit}}\sim 100$, and $r\sim 10$ gives
$D_{\mathrm{gate}}\sim 10^{11}$. With $\delta\sim 10^{-3}$, this yields the
conservative bound
\[
p \lesssim 10^{-14}.
\]
This estimate is highly conservative. Using optimized Hamiltonian
factorizations, such as integral tensor factorization~\cite{Huggins2021Meas}, can reduce $L$ to
$\mathrm{O}(n_{\mathrm{qubit}})$, leading instead to a weaker bound
\[
p \lesssim 10^{-8}.
\]

\twocolumngrid

\bibliographystyle{apsrev4-2}
\bibliography{main}

\end{document}